\documentclass[
 reprint,
 amssymb
]{revtex4-2}

\usepackage{amsthm}
\usepackage{hyperref}
\usepackage{mathtools}
\usepackage{graphicx}

\graphicspath{ {.} }
\newtheorem{theorem}{Theorem}

\begin{document}

\title{Full time-dependent counting statistics of highly entangled biphoton
states}
\author{Julian K. Nauth}
\email{nauth@posteo.de}
\affiliation{Institut f\"{u}r Angewandte Physik, Technical University of
Darmstadt, D-64289 Darmstadt, Germany}

\begin{abstract}
\noindent
Highly entangled biphoton states, generated by spontaneous parametric processes,
find wide applications in many experimental realizations. There is an increasing
demand for accurate prediction of their time-dependent detection. Unlike
approaches that have emerged so far, this paper presents an approach providing
full time-dependent counting statistics in terms of efficiently computable
formulas, valid for a wide range of entanglement and arbitrary interaction 
times. General spatial modes are taken into account to describe free space and
fiber propagation. The time intervals that correspond to the statistics are 
classified according to their widths. Apart from large and small widths compared
to the temporal correlation width, intermediate interval widths give access to
accidental correlations between separated time intervals. Moreover, the approach
is easily applicable to a modular array of arbitrary optical components and
external influences. This is demonstrated on phase-time coding, where the
detuning of the interferometers affecting Franson interference is investigated.
An acceptable range for the detuning is estimated, such that the security of the
key is not compromised.
\end{abstract}

\maketitle
\section{Introduction}
Quantum entanglement is one of the most useful resources in quantum information
technologies and is fundamental for secure information processing and
communications. In particular, a high degree of entanglement is attractive for
its high data capacity and error resilience \cite{Xie_2015}. Bipartite states
with continuous variables (CV) enable a degree of entanglement, which might be
much higher than the maximal achievable degree of entanglement with respect to
(w.r.t.) discrete variables \cite{Mikhailova_2008}. One of the most conventional
methods in CV is presently the generation of entangled photons by spontaneous
parametric processes, such as spontaneous parametric down-conversion (SPDC) and
spontaneous four-wave mixing (SFWM). These processes generate photon pairs with
a strong spectral and temporal correlation, where each photon is transmitted to
a party (conventionally referred to as Alice and Bob). It plays a major role in
quantum cryptography (QC) since the detection of one photon heralds the presence
of the other. Highly entangled biphoton states find a wide application in many
experimental realizations. Frequency-bin entanglement \cite{Imany_2018},
frequency multiplexing \cite{Joshi_2018}, nonlocal dispersion cancellation (NDC)
\cite{Li_2019}, and quantum electro-optic circuits \cite{Luo_2020} were recently
investigated using a continuous-wave pump laser. Although perfect frequency
entanglement can thus be achieved \cite{Zhai_2017}, it is inadequate for many
applications requiring timing information since the emission time of the photon
pairs is completely random \cite{Quan_2015}. Hence, pulsed pumps with
well-defined emission times were recently realized to generate highly entangled
photon pairs, where entanglement properties \cite{Quan_2015} and NDC
\cite{Xiang_2020,Lerch_2017,Hu_2015} have been investigated. Pulsed pumps have
also found great applications in areas such as arrayed waveguide grating
\cite{Matsuda_2017}, on-chip generation \cite{Kues_2017}, time-bin
implementation \cite{Martin_2017}, and quantum key distribution systems
\cite{Lee_2014,Arahira_2016}.

A specific application of pulsed pumps is phase-time coding, which was first
proposed by Brendel et al. \cite{Brendel_1998} and has been experimentally
realized \cite{Stucki_2006}. This is a promising approach for QC since coding in
time basis is particularly stable and the coherence length of the pump laser is
not critical \cite{Gisin_2001}. The setup includes a Franson arrangement, which
comprises two unbalanced Mach-Zehnder interferometers (MZIs) and is widely used
in several applications, such as gate operations \cite{Lo_2020}, superdense
coding \cite{Pavicic_2016}, and chip-based microresonators \cite{Peacock_2016}.
The key bits are first established based on the photon pair's temporal
correlation. Hence, there is a strong demand for an accurate prediction of this
correlation with ever-increasing precision \cite{Mueller_2020}. A second basis
to establish the key bits is implemented using Franson interference, which can
be achieved by adjusting the phases of the MZIs. To ensure perfect correlation,
the optical path lengths of the MZIs must differ less than the coherence length
of the entangled photons \cite{Gao_2019}. Otherwise, Franson interference is
affected, culminating in its complete disappearance if the detuning of the MZIs
exceeds the coherence length. Since the security of the key relies on perfect
correlation, full knowledge of this influence is indispensable. In particular,
an acceptable range of the detuning without compromising the security of the key
is of substantial interest for experimental realizations of phase-time coding.
For a detailed description, however, many kinds of temporal information must be
taken into account, such as the width of the pulsed pump, required for coding in
time basis, and the detailed temporal correlation, which determines Franson
interference in the case of small detuning. At the same time, a precise
description of detecting multiple photon pairs should be provided since
uncorrelated photons cause accidental coincidences, which must be discarded
during key distillation \cite{Takesue_2010,Gisin_2001}. Due to wide applications
and decisive advantages of high entanglement, an arbitrarily high degree of
entanglement should also be allowed, in particular the limit of perfect
entanglement.

A detailed description of Franson interference for small detuning, however, has
only been poorly addressed yet. Although many different approaches have emerged
so far, none of them is suitable for this task since they do not fulfill all
aforementioned requirements at the same time. For instance, modeling each pulse
as orthogonal states
\cite{Yan_2021,Lo_2020,Ikuta_2016,Martin_2017,Takeoka_2015,Richart_2012}
neglects many kinds of temporal information. Another way in which they are taken
into account is the perturbative limit of low mean photon number
\cite{Zhai_2017,Hou_2016,Quan_2015,Xie_2015,Phehlukwayo_2020,Xiang_2020,
Liu_2020,Dorfman_2021}.
Although this approach was recently used to investigate Franson interference for
increasing detuning of the MZIs \cite{Lukens_2013,Zhang_2021}, the generation of
multiple photon pairs is neglected. This, however, is of particular importance
for phase-time coding due to serious consequences on the security of the key
\cite{Gisin_2001}.

Another approach is to assume perfect entanglement \cite{Kuklewicz_2005} or to
approximate the strong temporal and spectral correlation, respectively, by a
diagonal matrix of discrete bins \cite{Wang_2020,Kues_2017}. This approach has
been applied to investigate spatial quantum correlations \cite{Brambilla_2004}
and to derive full counting statistics \cite{Signorini_2020,Takesue_2010}. The
detailed temporal correlation between entangled photons, however, is neglected.
This is incompatible with modeling of Franson interference for small detuning of
the MZIs. Further approaches, such as Schmidt decomposition \cite{Mauerer_2009}
and general optical modes \cite{Burenkov_2017}, provide full knowledge of the
quantum state \cite{Lamata_2005,Law_2004}. Although the latter approaches
include all physical information, even pump field depletion \cite{Perina_2016},
the computational effort increases significantly for higher entanglement since
more modes contribute, hence an exceedingly time and resource consuming
approach. Many experimental realizations can thus not be modeled suitably since
improving the degree of entanglement is of broad interest. In particular, the
limit of perfect entanglement remains an open problem.

In this paper, a theoretical approach is presented providing full
time-dependent counting statistics of photon pairs generated by spontaneous 
parametric processes in terms of efficiently computable formulas. The joint
amplitude may be chosen arbitrarily. General spatial modes and two widely
adopted communication channels, free space and fiber propagation \cite{Xu_2020},
are discussed. For the latter channel, the corresponding time intervals are
classified into two types: Their widths are much smaller and their widths are
much larger than the temporal correlation width. The first case provides the
computation of general correlation functions, whereas the second case is
important for coding in time basis. Intermediate interval widths are
investigated by expanding the latter liming cases, giving access to a more
precise prediction of accidental correlations between separated time intervals.
The approach is valid for a high degree of entanglement and likewise for a low
degree if the mean photon number is bounded by $\langle\hat{N}\rangle\ll N_
\text{max}$. The higher the entanglement, the larger $N_\text{max}$. In the case
of low entanglement, the mean photon number is restricted by $N_\text{max}
\approx 1$, which corresponds to the perturbative limit. In the case of high
entanglement, the limitation can be neglected due to $N_\text{max}\gg 1$. The
first steps of this approach were derived in a previous work
\cite{Master-Thesis} and the present paper shows it completely.

In the case of further optical components and external influences, the approach
can be generalized by simple modifications of the derived formulas. To this end,
a general procedure is presented, which can be easily applied to a modular array
of arbitrary components and influences. This is applied to the setup of
phase-time coding. To demonstrate the usefulness of this approach,
full counting statistics of Franson interference are derived and detection
probabilities are presented for increasing detuning of the MZIs. By means of
these results, an acceptable range for the detuning is estimated, such that the
security for practical QKD installations is not compromised. This could not be
achieved before since, unlike previous approaches, the requirements to
accurately describe Franson interference for increasing detuning are fulfilled.
All temporal information and multiple photon pairs effects are taken into
account. Since the derived formulas are efficiently computable, they can be
easily applied to an arbitrarily high degree of entanglement.

The paper is organized as follows: After the description of biphoton states
generated by spontaneous parametric processes in Sec. \ref{s:SPDC}, the general
photoelectric counting statistics are presented in Sec. \ref{s:statistics}. Even
though there is no systematic numerical treatment of the general statistics,
Sec. \ref{s:results} provides the statistics in terms of efficiently computable
formulas, which are the central results of this paper. At the very heart of
the derivation lies a theorem to evaluate the operators occurring in the general
statistics. In Sec. \ref{s:setup,influences}, a general procedure to generalize
the derived formulas to setups and external influences is presented. This is
applied to phase-time coding and the detuning of the MZIs affecting Franson
interference is investigated. Section \ref{s:conclusion} concludes the paper by
summarizing the main results.

\section{Biphoton states}\label{s:SPDC}
The approach of this paper deals with squeezed biphoton states, which are
generated by spontaneous parametric processes under certain assumptions. The
pump pulse is assumed to be bright enough to apply the undepleted pump
approximation, where appreciable attenuation due to down-conversion events can
be neglected \cite{Schneeloch_2018}. To treat frequencies continuously, the
crystal is considered to be large compared to the optical wavelengths
\cite{Schneeloch_2018}. For the sake of simplicity, the polarizations of input
and output photons are considered to be fixed, determining the type of the
process. Although omitted here, the results of this paper can be easily
extended to general polarization. Within these assumptions and the usual
rotating wave approximation, the Hamiltonian of SPDC processes in the
interaction picture reads as \cite{Schneeloch_2018,Ou_2007}
\begin{align}
\notag
\hat{H}(t) \propto&
\int_{\mathbb{R}^3} \text{d}^3 k_\text{A} \, \text{d}^3 k_\text{B} \, \text{d}^3
k_\text{P}\, \hat{a}^\dagger_\text{A}(\boldsymbol{k}_\text{A})
\hat{a}^\dagger_\text{B}(\boldsymbol{k}_\text{B})
\\\notag&
\times F(\boldsymbol{k}_\text{A},\boldsymbol{k}_\text{B},
\boldsymbol{k}_\text{P}) e^{i(\omega_\text{A}+\omega_\text{B}-\omega_\text{P})t} 
+ \text{H.c.},
\\\notag
F(\boldsymbol{k}_\text{A},\boldsymbol{k}_\text{B},\boldsymbol{k}_\text{P})
\coloneqq&\, \alpha_k(\boldsymbol{k}_\text{P}) \int_V \text{d}^3 r\, \chi^{(2)}
(\boldsymbol{r},\omega_\text{A},\omega_\text{B},\omega_\text{P})
\\&\times
e^{-i(\boldsymbol{k}_\text{A}+\boldsymbol{k}_\text{B}-\boldsymbol{k}_\text{P})
\cdot\boldsymbol{r}},
\end{align}
where $V$, $\alpha_k$, and $\chi^{(2)}$ denote the crystal volume, the momentum
amplitude of the pump field, and the nonlinear susceptibility. The frequencies
$\omega_i$ are determined by the corresponding momenta $\boldsymbol{k}_i$. The
mean momenta of both parties can be neglected if the photons are not mixed and
the propagation is devoid of nonlinear dispersion effects. Hence, they will be
set to zero, as well as the mean frequencies of fiber coupled photon pairs. To
obtain the Hamiltonian of the SFWM process, it suffices to make the substitution
\cite{Garay-Palmett_2013}
\begin{align}
\notag
F(\boldsymbol{k}_\text{A},\boldsymbol{k}_\text{B},\boldsymbol{k}_\text{P})
\coloneqq&\, \int_{\mathbb{R}^3} \text{d}^3 k'
\alpha_{k 1}(\boldsymbol{k}')\alpha_{k 2}(\boldsymbol{k}_\text{P}-
\boldsymbol{k}') \int_V \text{d}^3 r\,
\\& \chi^{(2)}(\boldsymbol{r},\omega_\text{A},\omega_\text{B},\omega')e^{-
i(\boldsymbol{k}_\text{A}+\boldsymbol{k}_\text{B}-\boldsymbol{k}')\cdot
\boldsymbol{r}}
\end{align}
with the corresponding pump fields $\alpha_{k 1}$ and $\alpha_{k 2}$. Due to the
dependence $\chi^{(2)}(\boldsymbol{r})$, techniques such as quasi-phase-matching
are included in this approach
\cite{Hu_2015,Yang_2016,Schneeloch_2019,Luo_2020,Signorini_2020}. An expression
of the time evolution operator for general interaction times $t$ can be achieved
using the Magnus expansion \cite{Quesada_2014,Blanes_2008}
\begin{align}
\notag
\hat{U} = &\exp\left[
\frac{1}{i\hbar}\int_{-t/2}^{t/2}\text{d}t'\, \hat{H}(t')
\right.\\&\left.-
\frac{1}{2\hbar^2}\int_{-t/2}^{t/2}\text{d}t'\int_{-t/2}^{t'}\text{d}t''\, 
[\hat{H}(t'),\hat{H}(t'')] + \ldots
\right].
\end{align}
Since $[\hat{a}_i \hat{a}_j,\hat{a}^\dagger_i \hat{a}^\dagger_j,
\hat{a}^\dagger_i \hat{a}_j,\mathbb{I}]_{i,j}$ forms a Lie algebra, $\hat{U}$
can be disentangled to a rotation operator $\hat{R}$ and a squeezing operator
\begin{align}\label{eq:S_op}
\hat{S} = \exp\left[
\int_{\mathbb{R}^3} \text{d}^3 k_\text{A} \, \text{d}^3 k_\text{B} \,
\psi(\boldsymbol{k}_\text{A},\boldsymbol{k}_\text{B})
\hat{a}^\dagger_\text{A}(\boldsymbol{k}_\text{A}) \hat{a}^\dagger_\text{B}
(\boldsymbol{k}_\text{B}) - \text{H.c.}
\right],
\end{align}
where $\psi(\boldsymbol{k}_\text{A},\boldsymbol{k}_\text{B})$ designates the
joint amplitude (JA). Initial vacuum states of Alice and Bob eliminate $\hat{R}$
to give
\begin{align}
\hat{U}\vert 0\rangle_\text{A}\vert 0\rangle_\text{B}
= \hat{S}\hat{R}\vert 0\rangle_\text{A}\vert 0\rangle_\text{B}
= \hat{S}\vert 0\rangle_\text{A}\vert 0\rangle_\text{B},
\end{align}
which gives proof that for each order of the Magnus expansion, even for the
exact solution, the quantum state can be written in terms of a squeezing
operator $\hat{S}$, determined by a JA $\psi(\boldsymbol{k}_\text{A},
\boldsymbol{k}_\text{B})$ as in Eq. (\ref{eq:S_op}). For the first-order term of
the Magnus expansion, the JA is given by
\begin{align}\label{eq:ja_magnus}
\notag
\psi(\boldsymbol{k}_\text{A},\boldsymbol{k}_\text{B}) \propto&
\int_{\mathbb{R}^3} \text{d}^3 k_\text{P} F(\boldsymbol{k}_\text{A},
\boldsymbol{k}_\text{B},\boldsymbol{k}_\text{P})
\\&
\times t\, \text{sinc}\left[t\frac{\omega_\text{A}+\omega_\text{B}-
\omega_\text{P}}{2}\right],
\end{align}
where time ordering effects are neglected and the limit of long interaction
times results in energy conservation. Further orders of the Magnus expansion are
discussed in \cite{Quesada_2014}, where an explicit expression of
$\psi(\boldsymbol{k}_\text{A},\boldsymbol{k}_\text{B})$ is derived in dependence
on $F(\boldsymbol{k}_\text{A},\boldsymbol{k}_\text{B},\boldsymbol{k}_\text{P})$.

Despite the closed-form expression of the quantum state, the continuous
character of the JA renders it difficult to compute the state exactly.
Therefore, many approximations pertaining to the aforementioned approaches, such
as the perturbative limit, perfect entanglement, and Schmidt decomposition, have
been applied to Eq. (\ref{eq:S_op}). In this manner, the approach of this
paper is based on a suitable expression of the JA, aiming to apply
approximations that are appropriate for highly entangled biphoton states and
still lead to analytical expressions for the counting statistics. To this end,
it is worthwhile to distill crucial features of high entanglement and to
correspondingly rewrite the JA $\psi(\boldsymbol{k}_\text{A},
\boldsymbol{k}_\text{B})$. First, it should be noted that it depends
distinctly on $\boldsymbol{k}_\text{A}+\boldsymbol{k}_\text{B}$ and
$\omega_\text{A}+\omega_\text{B}$ and thus exhibits a diagonal shape. Therefore,
it is plausible to express the JA in rotating coordinates
$\boldsymbol{k}_\text{A}+\boldsymbol{k}_\text{B}$ and
$\boldsymbol{k}_\text{A}-\boldsymbol{k}_\text{B}$. Secondly, the amplitudes
along the rotating coordinates determine spatial and momentum detection of each
party's pulse. This motivates the definition of the function
$\phi(\boldsymbol{\chi},\boldsymbol{\kappa})$, where normalized (dimensionless)
spatial and momentum coordinates $\boldsymbol{\chi}$ and $\boldsymbol{\kappa}$
represent the spatial and momentum amplitude, respectively, of each party's
pulse. As derived in Sec. \ref{s:results}, spatial and momentum detection can
simply be obtained by integration w.r.t. the other variable, similar to the
Wigner function. It should be pointed out that $\phi(\boldsymbol{\chi},
\boldsymbol{\kappa})$ is only assumed to have normalized widths in all variables
and otherwise can be chosen arbitrarily. In particular, $\phi(\boldsymbol{\chi},
\boldsymbol{\kappa})$ does not need to decouple, which allows the approach of
this paper to include non-trivial phase mismatches.

It is envisioned that the sought-after expression for the JA explicitly
reproduces each party's spatial and momentum amplitudes in its respective basis.
The JA $\psi(\boldsymbol{k}_\text{A},\boldsymbol{k}_\text{B})$ in momentum basis
and $\psi(\boldsymbol{x}_\text{A},\boldsymbol{x}_\text{B})$ in spatial basis can
be related by inserting
\begin{align}
\hat{a}(\boldsymbol{x}) &= \frac{1}{\sqrt{2\pi}^3} \int_{\mathbb{R}^3}
\text{d}^3 k\, \hat{a}(\boldsymbol{k}) e^{-i\boldsymbol{x}\cdot\boldsymbol{k}}
\end{align}
in Eq. (\ref{eq:S_op}). Due to the scaling property of the Fourier
transformation, it is apparent that high momentum anti-correlation corresponds
to high spatial correlation and vice versa. In the case of momentum
anti-correlation, each party's spatial and momentum amplitude are thus
determined by the JA along $\boldsymbol{x}_\text{A}-\boldsymbol{x}_\text{B}$ and
$\boldsymbol{k}_\text{A}+\boldsymbol{k}_\text{B}$, respectively. The need to
relate these coordinates to the normalized variables $\boldsymbol{\chi}$ and
$\boldsymbol{\kappa}$ suggests the introduction of three-dimensional momentum
and spatial widths. This can be written as
\begin{align}
\boldsymbol{\chi} = \frac{\Delta_{\boldsymbol{x}}^{-1}}{2}
(\boldsymbol{x}_\text{A}+\boldsymbol{x}_\text{B}) \;,\;
\boldsymbol{\kappa} = \frac{\delta_{\boldsymbol{x}}}{2}(\boldsymbol{k}_\text{A}-
\boldsymbol{k}_\text{B})
\end{align}
with positive-definite $3\times 3$ matrices $\delta_{\boldsymbol{x}}$ and
$\Delta_{\boldsymbol{x}}$, the eigenvalues of which designate the spatial 
correlation widths and the spatial full widths at half maximum (FWHM) of the
pulses at each party's side, respectively. Since $\delta_{\boldsymbol{x}}$ and
$\Delta_{\boldsymbol{x}}$ do not need to be diagonal, the approach of this
paper includes general elliptical shapes of the JA, essential for
non-collinear geometries \cite{Molina-Terriza_2005}.

In keeping with this modeling, the relation between the JA and
$\phi(\boldsymbol{\chi},\boldsymbol{\kappa})$ is set to be
\begin{align}\label{eq:ja}
\notag
\psi(\boldsymbol{k}_\text{A},&\boldsymbol{k}_\text{B})
= \det\left(\frac{\Delta_{\boldsymbol{x}}}{\sqrt{2\pi}}\right)
\\\notag&\times
\mathcal{F}^{-1}_{\boldsymbol{\chi}}
\left[\phi\left(\boldsymbol{\chi},\frac{\delta_{\boldsymbol{x}}}{2}
(\boldsymbol{k}_\text{A}-\boldsymbol{k}_\text{B})\right) \right]
\bm{(}\Delta_{\boldsymbol{x}}
(\boldsymbol{k}_\text{A}+\boldsymbol{k}_\text{B})\bm{)},
\\[0.5em]\notag
\psi(\boldsymbol{x}_\text{A},&\boldsymbol{x}_\text{B})
= \det \left(\frac{\delta_{\boldsymbol{x}}^{-1}}{\sqrt{2\pi}}\right)
\\&\times
\mathcal{F}_{\boldsymbol{\kappa}}
\left[\phi\left(\frac{\Delta_{\boldsymbol{x}}^{-1}}{2}
(\boldsymbol{x}_\text{A}+\boldsymbol{x}_\text{B}),
\boldsymbol{\kappa}\right) \right]
\bm{(}\delta^{-1}_{\boldsymbol{x}}
(\boldsymbol{x}_\text{A}-\boldsymbol{x}_\text{B})\bm{)},
\end{align}
where $\mathcal{F}$ denotes the Fourier transformation w.r.t. the subscripted
variable. It becomes apparent that these expressions for
$\psi(\boldsymbol{k}_\text{A},\boldsymbol{k}_\text{B})$ and
$\psi(\boldsymbol{x}_\text{A},\boldsymbol{x}_\text{B})$ indeed represent the
respective amplitudes of each party's pulse. The prefactors are chosen in such a
way that the structure of Eq. (\ref{eq:ja}) will be preserved at best, as
presented in Sec. \ref{s:results}. After the photon pairs have been coupled into
fibers, it suffices to consider the temporal correlation width $\delta_t$ and
the temporal FWHM $\Delta_t$ of each party's pulse, defined in Sec.
\ref{s:results}, where fiber coupling is discussed. It should be kept in mind
that $\delta_t$ and $\Delta_t$ are scalars since fiber coupling reduces the
spatial and momentum modes to one dimension.

In the case of collinear degenerate phase matching, where transverse momenta are
negligible, the JA in Eq. (\ref{eq:ja}) can be expressed in time and frequency
basis as
\begin{align}\label{eq:ja_t}
\notag
\psi(\omega_\text{A},&\omega_\text{B})
= \frac{\Delta_t}{\sqrt{2\pi}}
\\\notag&
\times\mathcal{F}^{-1}_{\chi}\left[\phi\left(\chi,
\delta_t\frac{\omega_\text{A}-\omega_\text{B}}{2}\right) \right]
\bm{(}\Delta_t(\omega_\text{A}+\omega_\text{B})\bm{)},
\\[0.5em]
\psi(t_\text{A},t_\text{B}) &= \frac{1}{\sqrt{2\pi}\delta_t}
\mathcal{F}_{\kappa}\left[\phi\left(\frac{t_\text{A}+t_\text{B}}{2\Delta_t},
\kappa\right) \right]\left(\frac{t_\text{A}-t_\text{B}}{\delta_t}\right),
\end{align}
where $\phi(\chi,\kappa)$ is intended to represent the temporal and spectral
amplitude of each party's pulse and, once again, is only assumed to have
normalized widths. All results of this paper can be reduced to this case by
replacing the matrices $\delta_{\boldsymbol{x}}$ and $\Delta_{\boldsymbol{x}}$
by the scalars $\delta_t$ and $\Delta_t$, respectively. Needless to say,
determinants of $\delta_t$ and $\Delta_t$ can be dropped.

Quantitative results will be computed for type-II SPDC with collinear degenerate
phase matching and long interaction time, neglecting time ordering effects,
where
\begin{align}\label{eq:ja_magnus_lim}
\psi(\omega_\text{A},\omega_\text{B}) \propto&\; F(\omega_\text{A},
\omega_\text{B},\omega_\text{A}+\omega_\text{B}).
\end{align}
Moreover, the phase mismatch is assumed to be dominated by
$\omega_\text{A}-\omega_\text{B}$, whereas
$\omega_\text{A}+\omega_\text{B}\approx  0$ compared to $\alpha_\omega$. This
enables to decouple the function $\phi(\chi,\kappa)$ as
\begin{align}\label{eq:phi_simple}
\phi(\chi,\kappa) \propto \alpha(\chi)\,\text{sinc}\left(\frac{\kappa}{2}
\right),
\end{align}
where the scale factor depends on the mean photon number and the degree of
entanglement, as discussed in Sec. \ref{s:results}. This determines the JA in
time basis to be
\begin{align}
\psi(t_\text{A},t_\text{B}) \propto \alpha\left(\frac{t_\text{A}+t_\text{B}}
{2\Delta_t}\right)\text{rect}\left(\frac{t_\text{A}-t_\text{B}}{\delta_t}
\right),
\end{align}
where the rectangular function rect$(x)$ originates from the crystal geometry.
The temporal pump amplitude $\alpha$ will be considered as Gaussian, where
$|\alpha|^2$ has normalized FWHM. It can be noted that $\delta_t$ and $\Delta_t$
correspond to their introduced interpretation.

In this paper, momentum anti-correlation and spatial correlation, that is,
$\Delta_{\boldsymbol{x}}>\delta_{\boldsymbol{x}}$, are discussed. In the
opposite case, the following results can be obtained in the same way, but this
might be discussed elsewhere. Conventionally, the degree of entanglement is
characterized by the Schmidt number $K$ \cite{Law_2004}, a measure for the
number of effectively contributing optical modes \cite{Mauerer_2009}. Another
important entanglement quantifier is given by the parameter $R$, defined as the
ratio of widths of single-particle and coincidence wave packets, which is rather
easily experimentally measurable \cite{Mikhailova_2008}. In three dimensions, the
latter corresponds to the matrix
$\Delta_{\boldsymbol{x}}\delta_{\boldsymbol{x}}^{-1}$, which is intended to
characterize the entanglement of the spatial (transverse and longitudinal) modes
in this paper. The Schmidt number $K$ turns out to be proportional to
$\det(\Delta_{\boldsymbol{x}}\delta_{\boldsymbol{x}}^{-1})$, as discussed in
Sec. \ref{s:results}. Entanglement of transverse modes is mainly determined by
the pump beam waist, in particular for large beam sizes
\cite{Molina-Terriza_2005}, and the transverse correlation length, recently
estimated in \cite{Schneeloch_2018}. Both quantities can be obtained by
projection of $\Delta_{\boldsymbol{x}}$ and $\delta_{\boldsymbol{x}}$ onto the
transverse plane, respectively. Various values of their ratio are used in
experimental realizations: 3.8 \cite{Hou_2016}, 13-79 \cite{Brambilla_2004},
and 140 \cite{Toninelli_2020} up to 360 \cite{Zhang_2008}.

For fiber coupled photon pairs, the degree of entanglement simplifies to
$\Delta_t/\delta_t$, which is particularly determined by the longitudinal modes.
Experimental realizations of entangled biphoton states deal with various values
of $\Delta_t/\delta_t$: 3.6 \cite{Matsuda_2017}, 5.2 \cite{Arahira_2016}, and 10
\cite{Ikuta_2016,Martin_2017} up to $10^3$ \cite{Richart_2012} and $10^5$
\cite{Hu_2015}. The approach of this paper is valid for a wide range of
entanglement, where $\Delta_{\boldsymbol{x}}\delta_{\boldsymbol{x}}^{-1}$ is
related to an upper bound of the mean photon number
$\langle\hat{N}\rangle\ll N_\text{max}$. The larger
$\Delta_{\boldsymbol{x}}\delta_{\boldsymbol{x}}^{-1}$, the larger
$N_\text{max}$, as discussed in Sec. \ref{s:results}.

\section{Photoelectric counting statistics}\label{s:statistics}
In this section, the photoelectric counting statistics of Alice and Bob are
addressed. Their photodetectors are first assumed to be perfect. As
communication channels, free space and fiber propagation are discussed, which
are both widely adopted \cite{Xu_2020}.

In the case of fiber propagation, the counting statistics of Alice and Bob are
referred to time intervals $I_\text{A},I_\text{B}$. The probability of $n$
counts at the time interval $I$ is given by \cite{MandelWolf_1995}
\begin{align}
p(n,I) =  \left\langle :\frac{\left(\hat{N}_I\right)^n}{n!}e^{-\hat{N}_I}:\right
\rangle
= \frac{1}{n!} g^{(n)}(0),
\end{align}
where $\hat{N}_I = \int_I \text{d}t \,\hat{a}^\dagger_t \hat{a}_t$ denotes the
photon number at $I$, the colons signify normal ordering, and
\begin{align}
g(y) &\coloneqq \left\langle : e^{-(1-y)\hat{N}_{I}}
:\right\rangle
= \left\langle  y^{\hat{N}_{I}}
\right\rangle
\end{align}
denotes the probability generating function (PGF) \cite{Johnson_2014}. This can
be extended to two parties as
\begin{align}\label{eq:g_def}
g(y_\text{A},y_\text{B}) &\coloneqq \left\langle  y_\text{A}^{
\hat{N}_{\text{A},I_\text{A}}}y_\text{B}^{\hat{N}_{\text{B},I_\text{B}}}
\right\rangle .
\end{align}
In the case of biphoton states generated by  spontaneous parametric processes,
this turns out to be
\begin{align}\label{eq:g_general}
\notag
g(y_\text{A},y_\text{B}) =& \det\left[ \mathbb{I} + P(y_\text{A},y_\text{B}) 
\left(\sigma_\text{f}-\frac{1}{2}\mathbb{I} \right) \right]^{-\frac{1}{2}} ,
\\[0.5em]
P(y_\text{A},y_\text{B}) \coloneqq& \begin{pmatrix}
(1-y_\text{A})P_{I_\text{A}} & 0 \\ 0 & (1-y_\text{B})P_{I_\text{B}}
\end{pmatrix},
\end{align}
where $\mathbb{I}$, $P_I$, and $\sigma_\text{f}$ are the identity operator, the
projection onto $I$ in time basis, and the covariance $\sigma$ of the biphoton
state projected onto the transverse modes that are selected by the fiber
\cite{Walborn_2010}, respectively. The derivation of Eq. (\ref{eq:g_general}) is
presented in Appendix \ref{app:statistics}. The Fredholm determinant included in
this result covers the correlations of all combinations of parties and time
values, except for time values outside of the intervals, which are eliminated by
the projection operators. It should be emphasized that Fredholm determinants
have no systematic numerical treatment in general \cite{Bornemann_2009}, which
renders an exact computation virtually impossible. In the case of free space
propagation, the counting statistics of Alice and Bob are referred to volumes
$V_\text{A},V_\text{B}\subseteq\mathbb{R}^3$. Similarly, the PGF is given by
\begin{align}
\notag
g(y_\text{A},y_\text{B}) =& \det\left[ \mathbb{I} + P(y_\text{A},y_\text{B}) 
\left(\sigma-\frac{1}{2}\mathbb{I} \right) \right]^{-\frac{1}{2}} , \\[0.5em]
P(y_\text{A},y_\text{B}) \coloneqq& \begin{pmatrix}
(1-y_\text{A})P_{V_\text{A}} & 0 \\ 0 & (1-y_\text{B})P_{V_\text{B}}
\end{pmatrix}
\end{align}
with the projection $P_V$ onto $V$ in spatial basis.

These results can be generalized to real photodetectors. Dark counts can be 
modeled by additional radiation with Poisson or thermal distributed counting
statistics \cite{Paris_2001}. Quantum efficiencies $\eta_\text{A},\eta_\text{B}$
can be taken into account by modifying the photon number operator
$\hat{N}_{q,I_q}\mapsto \eta_q\hat{N}_{q,I_q}$ of both parties
$q=\text{A},\text{B}$ \cite{Ferraro_2005}, which leads to an effective
replacement
\begin{align}\label{eq:g_ineff}
g(y_\text{A},y_\text{B}) \mapsto g(1-\eta_\text{A}+\eta_\text{A} y_\text{A},1-
\eta_\text{B}+\eta_\text{B} y_\text{B}).
\end{align}
It can be shown that attenuation in fibers leads to the same result, which can 
be modeled by beam splitters (BSs), where the reflected outcomes are eliminated 
\cite{Takeoka_2015}.

The covariance operator $\sigma$ of the biphoton state describes the correlation
between two arbitrary time values within each party and among both parties.
Expressed in terms of the JA
$\psi(\boldsymbol{x}_\text{A},\boldsymbol{x}_\text{B})$, given by Eq.
(\ref{eq:ja}), the covariance reads as \cite{MaRhodes_1990}
\begin{align}\label{eq:sigma}
\sigma =
\begin{pmatrix}
\sigma_\text{AA} & \sigma_\text{AB}
\\
\sigma_\text{BA} & \sigma_\text{BB}
\end{pmatrix} ,
\end{align}
where each entry can be written as
\begin{align}\label{eq:sigma_entry}
\notag
\sigma_\text{AB} = {\sigma_\text{BA}}^\text{T} &= \frac{1}{4}\begin{pmatrix}
su+\left(\tilde{s}\tilde{u}^\dagger\right)^\text{T} & -isu+i\left(\tilde{s}
\tilde{u}^\dagger\right)^\text{T}  \\
-isu +i\left(\tilde{s}\tilde{u}^\dagger\right)^\text{T} & -su-\left(\tilde{s}
\tilde{u}^\dagger\right)^\text{T}
\end{pmatrix},
\\\notag
\sigma_\text{AA} &= \frac{1}{4}\begin{pmatrix}
c+c^\text{T} & ic-ic^\text{T} \\
-ic+ic^\text{T} & c+c^\text{T}
\end{pmatrix},
\\
\sigma_\text{BB} &= \frac{1}{4}\begin{pmatrix}
\tilde{c}+\tilde{c}^\text{T} & i\tilde{c}-i\tilde{c}^\text{T} \\
-i\tilde{c}+i\tilde{c}^\text{T} & \tilde{c}+\tilde{c}^\text{T}
\end{pmatrix}
\end{align}
using the abbreviations $c\coloneqq\cosh(2r),\,\tilde{c}\coloneqq
\cosh(2\tilde{r}),\,{s}\coloneqq\sinh(2r),\, \tilde{s}\coloneqq
\sinh(2\tilde{r})$ and the polar decomposition of
\begin{align}\label{eq:r,u}
\psi = ru = \tilde{u}\tilde{r}
\end{align} 
in Hermitian parts $r,\tilde{r}$ and unitary parts $u,\tilde{u}$. It may be
worth noting that $\psi$ can be seen as an integral operator, where the JA
$\psi(\boldsymbol{x}_\text{A},\boldsymbol{x}_\text{B})$ determines the kernel.
As one might have anticipated, the squeezing process leads to hyperbolic
functions of the JA, occurring in Eq. (\ref{eq:sigma_entry}).

\section{Analytical results}\label{s:results}
A crucial problem for the quantification of general PGFs as in Eq.
(\ref{eq:g_general}) is the computation of the covariance $\sigma$. There is a
strong demand for a simple, analytical expression of the operators occurring in
Eq. (\ref{eq:sigma_entry}).

Before embarking on the crucial theorem of this paper providing a satisfactory
solution, the first, nonzero order of the PGF is addressed separately. For
$I_\text{A}=I_\text{B}=\mathbb{R}$, expanding the PGF in $r$ and $\tilde{r}$,
defined in Eq. (\ref{eq:r,u}), reveals the quadratic order, which can be
computed without any simplifications as
\begin{align}\label{eq:quadr_exact}
\text{Tr}\left[r^2\right] = \text{Tr}\left[\tilde{r}^2\right] = \det
\left(\frac{\Delta_{\boldsymbol{x}}\delta_{\boldsymbol{x}}^{-1}}{2\pi}\right) 
\Vert\phi\Vert_2^2,
\end{align}
where $\Vert .\Vert_p$ denotes the $L^p(\mathbb{R}^6)$-norm. It should be
emphasized that Eq. (\ref{eq:quadr_exact}) is devoid of any assumption of
$\Delta_{\boldsymbol{x}}$ and $\delta_{\boldsymbol{x}}$. Hence, the perturbative
limit $\langle\hat{N}\rangle\ll 1$ is covered by the following results for any
choice of $\Delta_{\boldsymbol{x}}$ and $\delta_{\boldsymbol{x}}$ whatsoever.
Moreover, this order can be bounded by the mean number of photons leaving the
crystal at one party's side (e.g., Alice) \cite{Johnson_2014}:
\begin{align}\label{eq:N_estim}
\langle\hat{N}\rangle &= \partial_\text{A}g(1,1) = \frac{1}{2}\text{Tr}\left[c-
\mathbb{I}\right]\geq
\text{Tr}[r^2],
\end{align}
where the PGF $g(y_\text{A},y_\text{B})$ is given by Eq. (\ref{eq:g_general}).
Provided that $\langle\hat{N}\rangle<\infty$, it can be seen that $r$ and
$\tilde{r}$ are Hilbert-Schmidt operators \cite{Bornemann_2009}. Under further
assumptions that are physically reasonable, $r$ and $\tilde{r}$ can even be
shown to be trace class operators \cite{Bornemann_2009}.

A detailed quantification of the PGFs should also include higher orders of $r$
and $\tilde{r}$. To this end, the expression of the JA, as introduced in Eq.
(\ref{eq:ja}), benefits the following theorem providing a point-wise evaluation
of $f(r)$ and $f(\tilde{r})$ in spatial and momentum basis. It reveals the major
advantage of Eq. (\ref{eq:ja}), which consists in the fact that its structure is
completely preserved and functions of the JA simplify to functions of
$\phi(\boldsymbol{\chi},\boldsymbol{\kappa})$.
\begin{widetext}
\begin{theorem}\label{theorem}
Let $r$, $\tilde{r}$, $u$, and $\tilde{u}$ be as defined in Eq. (\ref{eq:r,u})
and let $f(x)$ be an analytic function fulfilling $f(0)=0$. \\If the terms of
order $n\gtrsim n_{\lim}\coloneqq 2/\Vert\Delta_{\boldsymbol{x}}^{-1}
\delta_{\boldsymbol{x}}\Vert_\sigma$ of its power series can be neglected for
all
\begin{align}
|x|\leq x_{\max}\coloneqq \sqrt{\det(2\pi\Delta_{\boldsymbol{x}}^{-1}
\delta_{\boldsymbol{x}})\langle\hat{N}\rangle} \frac{\Vert\phi\Vert_\infty}
{\Vert\phi\Vert_2},
\end{align}
where $\Vert .\Vert_\infty$, $\Vert .\Vert_p$, and $\Vert .\Vert_\sigma$ are the
supremum norm, the $L^p(\mathbb{R}^6)$-norm, and the spectral norm,
respectively, the following holds:
\begin{align}
\notag
\left[f\left(r\right)\right](\boldsymbol{x},\boldsymbol{x}')=\left[f
\left(\tilde{r}\right)\right](\boldsymbol{x},\boldsymbol{x}')
&=
\det\left(\frac{\delta_{\boldsymbol{x}}^{-1}}{\sqrt{2\pi}}\right)
\mathcal{F}_{\boldsymbol{\kappa}} \left\lbrace f \left[\left| \phi
\left(\frac{\Delta_{\boldsymbol{x}}^{-1}}{2}(\boldsymbol{x}+\boldsymbol{x}'),
\boldsymbol{\kappa}\right) \right|\right] \right\rbrace \bm{(}
\delta^{-1}_{\boldsymbol{x}}(\boldsymbol{x}-\boldsymbol{x}')\bm{)},
\\[1em]\notag
\left[f\left(r\right)u\right](\boldsymbol{x},\boldsymbol{x}') &=
\det\left(\frac{\delta_{\boldsymbol{x}}^{-1}}{\sqrt{2\pi}}\right)
\mathcal{F}_{\boldsymbol{\kappa}} \left\lbrace \tilde{f} \left[\phi
\left(\frac{\Delta_{\boldsymbol{x}}^{-1}}{2}(\boldsymbol{x}+\boldsymbol{x}'),
\boldsymbol{\kappa}\right) \right]\right\rbrace \bm{(}
\delta^{-1}_{\boldsymbol{x}}(\boldsymbol{x}-\boldsymbol{x}')\bm{)},
\\[1em]\notag
\left[f\left(\tilde{r}\right)\tilde{u}^\dagger\right](\boldsymbol{x},
\boldsymbol{x}') &=
\det\left(\frac{\delta_{\boldsymbol{x}}^{-1}}{\sqrt{2\pi}}\right)
\mathcal{F}_{\boldsymbol{\kappa}} \left\lbrace \tilde{f} \left[ \phi
\left(\frac{\Delta_{\boldsymbol{x}}^{-1}}{2}(\boldsymbol{x}+\boldsymbol{x}'),
\boldsymbol{\kappa}\right)^* \right]\right\rbrace \bm{(}
\delta^{-1}_{\boldsymbol{x}}(\boldsymbol{x}-\boldsymbol{x}')\bm{)},
\\[1.5em]
\notag
\left[f\left(r\right)\right](\boldsymbol{k},\boldsymbol{k}') = \left[f
\left(\tilde{r}\right)\right](-\boldsymbol{k},-\boldsymbol{k}')
&=
\det\left(\frac{\Delta_{\boldsymbol{x}}}{\sqrt{2\pi}}\right)
\mathcal{F}^{-1}_{\boldsymbol{\chi}} \left\lbrace f \left[ \left| \phi
\left(\boldsymbol{\chi},\frac{\delta_{\boldsymbol{x}}}{2}(\boldsymbol{k}+
\boldsymbol{k}')\right) \right| \right]\right\rbrace \bm{(}
\Delta_{\boldsymbol{x}}(\boldsymbol{k}-\boldsymbol{k}')\bm{)},
\\[1em]\notag
\left[f\left(r\right)u\right](\boldsymbol{k},\boldsymbol{k}')
&= \det\left(\frac{\Delta_{\boldsymbol{x}}}{\sqrt{2\pi}}\right)
\mathcal{F}^{-1}_{\boldsymbol{\chi}} \left\lbrace \tilde{f} \left[ \phi
\left(\boldsymbol{\chi},\frac{\delta_{\boldsymbol{x}}}{2}(\boldsymbol{k}-
\boldsymbol{k}')\right) \right]\right\rbrace \bm{(}\Delta_{\boldsymbol{x}}
(\boldsymbol{k}+\boldsymbol{k}')\bm{)},
\\[1em]
\left[f\left(\tilde{r}\right)\tilde{u}^\dagger\right](\boldsymbol{k},
\boldsymbol{k}')
&= \det\left(\frac{\Delta_{\boldsymbol{x}}}{\sqrt{2\pi}}\right)
\mathcal{F}^{-1}_{\boldsymbol{\chi}} \left\lbrace \tilde{f} \left[ \phi
\left(\boldsymbol{\chi},\frac{\delta_{\boldsymbol{x}}}{2}(-\boldsymbol{k}+
\boldsymbol{k}')\right)^* \right]\right\rbrace \bm{(}\Delta_{\boldsymbol{x}}(-
\boldsymbol{k}-\boldsymbol{k}')\bm{)},
\end{align}
where $ \tilde{f}(x)\coloneqq f\left(\left|x\right|\right)x/|x|$ and
$\phi(\boldsymbol{\chi},\boldsymbol{\kappa})$ is defined in Eq. (\ref{eq:ja}).
\end{theorem}
\end{widetext}
The proof can be found in Appendix \ref{app:theorems}. Owing to this theorem,
the covariance $\sigma$ can be evaluated point-wise in spatial basis as
\begin{align}\label{eq:sigma_S}
\notag
&\left[\sigma-\frac{1}{2}\mathbb{I}\right](\boldsymbol{x},\boldsymbol{x}')
\\
&=\det
\left(\frac{\delta_{\boldsymbol{x}}^{-1}}{\sqrt{2\pi}}\right)
\mathcal{F}_{\boldsymbol{\kappa}}  \left[S\left(\frac{\Delta_{
\boldsymbol{x}}^{-1}}{2}(\boldsymbol{x}+\boldsymbol{x}'),\boldsymbol{\kappa}
\right) \right] \bm{(}\delta^{-1}_{\boldsymbol{x}}
(\boldsymbol{x}-\boldsymbol{x}')\bm{)}&
\end{align}
in terms of the matrix
\begin{align}\label{eq:S(x',x)}
S(\boldsymbol{\chi},\boldsymbol{\kappa}) =
\begin{pmatrix}
S_\text{AA}(\boldsymbol{\chi},\boldsymbol{\kappa}) & S_\text{AB}
(\boldsymbol{\chi},\boldsymbol{\kappa})
\\
S_\text{BA}(\boldsymbol{\chi},\boldsymbol{\kappa}) & S_\text{BB}
(\boldsymbol{\chi},\boldsymbol{\kappa})
\end{pmatrix} .
\end{align}
The entries are given by
\begin{align}
\notag
S_\text{AA}(\boldsymbol{\chi},\boldsymbol{\kappa}) = S_\text{BB}
(\boldsymbol{\chi},\boldsymbol{\kappa})
 =& \frac{1}{4}\sum_\pm \phi_\text{c}(\boldsymbol{\chi},\pm \boldsymbol{\kappa}) 
 \begin{pmatrix}
1 & \pm i \\ \mp i & 1
\end{pmatrix},
\\
S_\text{AB}(\boldsymbol{\chi},\boldsymbol{\kappa}) = S_\text{BA}
(\boldsymbol{\chi},\boldsymbol{\kappa})^*
 =& \frac{1}{4}\sum_\pm
\phi_\text{s}(\boldsymbol{\chi},\pm \boldsymbol{\kappa})^\pm
\begin{pmatrix}
1 & \mp i \\ \mp i & -1
\end{pmatrix},
\end{align}
where $z^\pm \coloneqq \Re z \pm i\Im z$ is intended to abbreviate
$z\in\mathbb{C}$ itself and the complex conjugate $z^*$, respectively. Moreover,
the abbreviations
\begin{align}\label{eq:phi_c,s}
\notag
\phi_\text{c}(\boldsymbol{\chi},\boldsymbol{\kappa})&\coloneqq \cosh\bm{(}2
|\phi(\boldsymbol{\chi},\boldsymbol{\kappa})|\bm{)}-1,
\\
\phi_\text{s}(\boldsymbol{\chi},\boldsymbol{\kappa})&\coloneqq \sinh\bm{(}2
|\phi(\boldsymbol{\chi},\boldsymbol{\kappa})|\bm{)}\frac{\phi(\boldsymbol{\chi},
\boldsymbol{\kappa})}{|\phi(\boldsymbol{\chi},\boldsymbol{\kappa})|}
\end{align}
are used throughout this paper, where $\phi(\boldsymbol{\chi},
\boldsymbol{\kappa})$ is defined in Eq. (\ref{eq:ja}). In the case of collinear
degenerate phase matching, $\phi_\text{c}(\chi,\kappa)$ and $\phi_\text{s}(\chi,
\kappa)$ are defined in the same way, where $\phi(\boldsymbol{\chi},
\boldsymbol{\kappa})$ needs to be replaced by $\phi(\chi,\kappa)$, defined in
Eq. (\ref{eq:ja_t}).

Further interesting quantities, such as the Schmidt number $K$
\cite{Mikhailova_2008}, can also be calculated as
\begin{align}\label{eq:K}
K = \frac{\text{Tr}[r^2]^2}{\text{Tr}[r^4]}
= \det\left(\frac{\Delta_{\boldsymbol{x}}\delta_{\boldsymbol{x}}^{-1}}{2\pi}
\right) \left( \frac{\Vert\phi\Vert_2}{\Vert\phi\Vert_4} \right)^4,
\end{align}
which reveals the aforementioned relation to $\Delta_{\boldsymbol{x}}
\delta_{\boldsymbol{x}}^{-1}$. In the case of Eq. (\ref{eq:phi_simple}) with
collinear degenerate phase matching, $K$ and $\Delta_t/\delta_t$ even share the
same magnitude:
\begin{align}
\frac{K}{\Delta_t/\delta_t} = \frac{1}{2\pi}\left( \frac{\Vert\phi\Vert_2}{\Vert
\phi\Vert_4} \right)^4
= \frac{3}{2\sqrt{2\pi\ln(2)}} \approx 0.72 .
\end{align}
Based on theorem \ref{theorem}, the spatial and momentum distribution of each
party's pulse can be obtained as
\begin{align}\label{eq:N_omega}
\notag
\langle\hat{N}_{\text{A}}(\boldsymbol{x})\rangle = \langle\hat{N}_{\text{B}}
(\boldsymbol{x})\rangle =&
\frac{1}{2}\det\left(\frac{\delta_{\boldsymbol{x}}^{-1}}{2\pi}\right) 
\int_{\mathbb{R}^3}\text{d}^3\kappa\, \phi_\text{c}(\Delta_{\boldsymbol{x}}^{-1} 
\boldsymbol{x},\boldsymbol{\kappa}),
\\
\langle\hat{N}_{\text{A}}(\boldsymbol{k})\rangle = \langle\hat{N}_{\text{B}}(-
\boldsymbol{k})\rangle =&
\frac{1}{2}\det\left(\frac{\Delta_{\boldsymbol{x}}}{2\pi}\right) 
\int_{\mathbb{R}^3}\text{d}^3\chi\, \phi_\text{c}(\boldsymbol{\chi},
\delta_{\boldsymbol{x}}\boldsymbol{k}).
\end{align}
These relations corroborate the interpretation of $\delta_{\boldsymbol{x}}$ and
$\Delta_{\boldsymbol{x}}$ as well as the motivation of introducing
$\phi(\boldsymbol{\chi},\boldsymbol{\kappa})$, namely representing the spatial
and momentum amplitude of each party's pulse. Integrating Eq. (\ref{eq:N_omega})
gives the mean photon number of one party as
\begin{align}\label{eq:N}
\langle\hat{N}\rangle = \frac{1}{2}\det\left(\frac{\Delta_{\boldsymbol{x}}
\delta_{\boldsymbol{x}}^{-1}}{2\pi}\right) \int_{\mathbb{R}^3}\text{d}^3\kappa\,
\text{d}^3\chi\, \phi_\text{c}(\boldsymbol{\chi},\boldsymbol{\kappa}).
\end{align}
It should be kept in mind that this relation serves to determine the scale
factor of $\phi(\boldsymbol{\chi},\boldsymbol{\kappa})$ in dependence on the
mean photon number and the degree of entanglement, as pointed out for Eq.
(\ref{eq:phi_simple}). This can be done numerically or by expanding
$\phi_\text{c}(\boldsymbol{\chi},\boldsymbol{\kappa})$ in terms of
$\phi(\boldsymbol{\chi},\boldsymbol{\kappa})$.

The condition of theorem \ref{theorem} reveals the regime of validity of the
approach that is presented in this paper: The higher the entanglement
$\Delta_{\boldsymbol{x}}\delta_{\boldsymbol{x}}^{-1}$ and the lower the mean
photon number $\langle\hat{N}\rangle$, the better is the condition fulfilled.
The required degree of entanglement is thus related to the mean photon number,
which can be expressed by an upper bound $\langle\hat{N}\rangle\ll
N_\text{max}$. The relation between the degree of entanglement and
$N_\text{max}$ is presented in Tab. \ref{tab:N_max} and evaluated in the case of
Eq. (\ref{eq:phi_simple}). To derive this relation, the condition of theorem
\ref{theorem} can be adapted to collinear degenerate phase matching as
\begin{align}
n_\text{lim}= \frac{2\Delta_t}{\delta_t}, \quad
x_\text{max}= \sqrt{2\pi\langle\hat{N}\rangle\frac{\delta_t}{\Delta_t}} 
\frac{\Vert\phi\Vert_\infty}{\Vert\phi\Vert_2}.
\end{align}
In view of Eq. (\ref{eq:phi_c,s}), the remainder of the power series of
$f(x)\coloneqq \cosh(2x)-1$ and $\sinh(2x)$ needs to be estimated.
The remainder of the order $n\gtrsim n_\text{lim}$ has thus to be much smaller
than the quadratic and linear order for all $\vert x\vert \leq x_\text{max}$,
respectively. This can be achieved by
\begin{align}
\frac{\cosh(2x_\text{max})}{n!}(2x_\text{max})^{n}
\ll \min\left\lbrace x_\text{max} , 2x_\text{max}^2 \right\rbrace .
\end{align}
Since $N_\text{max}\gg\langle\hat{N}\rangle$ exceeds the regime of validity, the
variable $\tilde{x}\gg x_\text{max}$ is introduced, which is defined as
\begin{align}
\sqrt{2\pi N_\text{max}\frac{\delta_t}{\Delta_t}} \frac{\Vert\phi\Vert_\infty}
{\Vert\phi\Vert_2} \coloneqq \tilde{x}
\end{align}
and fulfills
\begin{align}
\frac{\cosh(2\tilde{x})}{n_\text{lim}!}(2\tilde{x})^{n_\text{lim}}
= \min\left\lbrace \tilde{x} , 2\tilde{x}^2 \right\rbrace.
\end{align}
Computing $\tilde{x}$ for fixed $\Delta_t/\delta_t$ gives $N_\text{max}$, which
is presented in Tab. \ref{tab:N_max}. It should be pointed out that the
computation of the quadratic order is valid for any choice of $\Delta_t$ and
$\delta_t$, as shown in Eq. (\ref{eq:quadr_exact}).
\begin{table}[t!]
\centering
{\renewcommand{\arraystretch}{1.2}
\begin{tabular}{ccccccccccccccccccc}
\hline\hline
$\Delta_t/\delta_t$ && 2 && 3 && 4 && 5 && 7 && 10 && 15 && 20
\\ \hline
$N_\text{max}$ && 2.9 && 8.7 && 18 && 34 && 85 && 229 && 715 && 1619
\\\hline\hline
\end{tabular}
}
\caption{Upper bound $N_\text{max}$ of the mean photon number for given degree
of entanglement $\Delta_t/\delta_t$, such that the approach of this paper is
still valid. This relation was evaluated in the case of Eq.
(\ref{eq:phi_simple}).}
\label{tab:N_max}
\end{table}

Moreover, theorem \ref{theorem} paves the way towards a simple, analytical
expression of the general PGF in Eq. (\ref{eq:g_general}), which will be derived
for fiber propagation. Although omitted in this paper, the same results can be
easily obtained for free space propagation by extending the formulas to
three-dimensional spatial basis, which might be discussed elsewhere. In the case
of fiber propagation, the covariance needs to be modified due to fiber coupling
as
\begin{align}\label{eq:sigma_proj}
\notag
\left[\sigma_\text{f}-\frac{1}{2}\mathbb{I}\right](x,x')
=&
\int_{\mathbb{R}^3}\text{d}^3q\,\text{d}^3q' P_\text{f}(x,\boldsymbol{q})
P_\text{f}(x',\boldsymbol{q}')^\text{T}
\\&\otimes
\left[\sigma-\frac{1}{2}\mathbb{I}\right](\boldsymbol{q}, \boldsymbol{q}') 
,
\end{align}
occurring in Eq. (\ref{eq:g_general}), where $P_\text{f}$ denotes the projection
onto transverse modes of the fiber (e.g., Laguerre Gaussian modes or Hermite
Gaussian modes) \cite{Walborn_2010}. It may be worth noting that $P_\text{f}
(x,\boldsymbol{q})P_\text{f}(x',\boldsymbol{q}')^\text{T}$ is generally a matrix
representing all combinations of transverse modes selected by the fiber. This
extends $\sigma-\mathbb{I}/2$ to an additional mode structure, hence the tensor
product. More precisely, a continuous number of transverse modes simplifies to a
finite number. Equation (\ref{eq:sigma_proj}) can be expressed in time basis and
can be put in the form
\begin{align}\label{eq:sigma_fiber}
\left[\sigma_\text{f}-\frac{1}{2}\mathbb{I}\right](t,t') = \frac{1}{\sqrt{2\pi}
\delta_t} \mathcal{F}_\kappa\left[ S\left(\frac{t+t'}{2\Delta_t},\kappa\right) 
\right]
\left(\frac{t-t'}{\delta_t}\right)&,
\end{align}
which defines the matrix $S(\chi,\kappa)$ and determines $\delta_t$ and
$\Delta_t$ since $S(\chi,\kappa)$ is once again assumed to have normalized
widths. In the case of collinear degenerate phase matching, fiber coupling in
Eq. (\ref{eq:sigma_proj}) is redundant and $S(\chi,\kappa)$ coincides with Eq.
(\ref{eq:S(x',x)}) depending on $\phi(\chi,\kappa)$, defined in Eq.
(\ref{eq:ja_t}), instead of $\phi(\boldsymbol{\chi},\boldsymbol{\kappa})$.

As discussed in Sec. \ref{s:statistics}, the counting statistics w.r.t. fiber
propagation are referred to time intervals $I_\text{A},I_\text{B}$. An
analytical expression of the general PGF, however, can hardly be achieved for
general $I_\text{A},I_\text{B}$. Therefore, they are classified according to
their widths into two types: $|I_\text{A}|,|I_\text{B}|\ll\delta_t$ and
$|I_\text{A}|,|I_\text{B}|\gg\delta_t$, discussed in Secs. \ref{s:spdc_small}
and \ref{s:spdc_large}, respectively. The first case provides the computation of
general correlation functions, whereas the second case is important for coding
in time basis. The remainder of this section addresses intermediate interval
widths, paving the way towards a more precise prediction of accidental
correlations between separated time intervals. The following results can also be
obtained in frequency basis, that is, $I_\text{A},I_\text{B}$ would designate
frequency intervals, but this is omitted here and might be discussed elsewhere.

\subsection{Small interval widths}\label{s:spdc_small}
The intervals are now assumed to satisfy $|I_\text{A}|,|I_\text{B}|\ll\delta_t$.
In order to facilitate the PGF in Eq. (\ref{eq:g_general}), the projection
operators $P_{I_\text{A}},P_{I_\text{B}}$ and the covariance $\sigma_\text{f}$
are addressed. According to Eq. (\ref{eq:sigma_fiber}), $\sigma_\text{f}$ varies
at most in the magnitude of $\delta_t$. Therefore, $P_{I_\text{A}}$ and
$P_{I_\text{B}}$ cause approximately a point-wise evaluation at some arbitrary
value in the interval. This can be written as
\begin{align}\label{eq:pointwise P_I}
\notag
P_{I_q} &\approx |I_q\rangle\langle I_q|, \\
\langle I_q|A|I_{q'}\rangle &\approx \sqrt{|I_{q}||I_{q'}|} A(T_{q},T_{q'})
\end{align}
for any operator $A= c,\tilde{c},s,\tilde{s}$ occurring in Eq.
(\ref{eq:sigma_entry}), and party $q$, where $T_q\in I_q$ is an arbitrary time
value and $\langle t|I_q\rangle\coloneqq \textbf{1}_{t\in I}/\sqrt{|I_q|}$
denotes the $L^2$-normalized indicator function of $I_q$. Inserting Eq.
(\ref{eq:sigma_fiber}) into the PGF gives
\begin{align}\label{eq:g_small}
\notag
g(y_\text{A},y_\text{B})
=&\det\left[ \mathbb{I}-\frac{1}{2\pi\delta_t}
\begin{pmatrix}
\tau_\text{A}(y_\text{A}) & 0 \\ 0 & \tau_\text{B}(y_\text{B})
\end{pmatrix}
\right.\\\notag\times&\left.
\int_{-\infty}^\infty\text{d}\kappa\,
\begin{pmatrix}
\tilde{S}_\text{AA}(T_\text{A},T_\text{A}) & \tilde{S}_\text{AB}(T_\text{A},
T_\text{B}) \\ \tilde{S}_\text{BA}(T_\text{B},T_\text{A}) & \tilde{S}_\text{BB}
(T_\text{B},T_\text{B})
\end{pmatrix}
\right]^{-\frac{1}{2}} ,
\\[0.5em]
\tilde{S}(T,T')\coloneqq & S\left( \frac{T+T'}{2\Delta_t},\kappa \right) e^{-i
\kappa(T-T')/{\delta_t}},
\end{align}
where $\tau_q(y_q) \coloneqq (y_q-1)|I_q|$ and $T_q\in I_q$ can be chosen
arbitrarily for each party $q$. It should be noted that Eq. (\ref{eq:g_small})
can be evaluated with moderate computational effort. Due to the reduction to one
time value per interval, only the correlation between the parties is taken into
account. The Fredholm determinant thus shrinks to a determinant of a $4\times 4$
matrix. Moreover, the Fourier transform, occurring in the entries AB and BA, can
be computed efficiently by fast Fourier transformation. Unlike approaches based
on Schmidt decomposition or general optical modes, the computation of the
determinant does not grow indefinitely for increasing entanglement.

Equation (\ref{eq:g_small}) serves to generalize the counting statistics to
further components and external influences, as discussed in Sec.
\ref{s:General Procedure}, and provides the computation of correlation functions
\begin{align}\label{eq:corr_general}
\langle \hat{N}_\text{A}(T_{\text{A}1})\ldots \hat{N}_\text{A}(T_{\text{A}
m_\text{A}})
\hat{N}_\text{B}(T_{\text{B}1})\ldots \hat{N}_\text{B}(T_{\text{B}m_\text{B}})
\rangle
\end{align}
of general order $m_\text{A},m_\text{B}\in\mathbb{N}$. To relate correlation
functions to general PGFs, defined in Eq. (\ref{eq:g_def}), it is pertinent to
replace the photon number operators as
\begin{align}\label{eq:N(T)_derive}
\hat{N}_q(T_q) = \partial_{|I_q|}  \partial_{y_q} \left( y_q^{\hat{N}_{q,I_q}} 
\right|
_{y_q=1,|I_q|=0},
\end{align}
where $I_q=[T_q-|I_q|/2,T_q+|I_q|/2]$ for each party $q$. Since the PGF in Eq.
(\ref{eq:g_small}) only depends on $\tau_\text{A}$ and $\tau_\text{B}$, the
derivatives in Eq. (\ref{eq:N(T)_derive}) simplify to $\partial_{\tau_q}$
evaluated at $\tau_q=0$. The second-order correlation function can thus be
written as
\begin{align}
\langle \hat{N}_\text{A}(T_{\text{A}})
\hat{N}_\text{B}(T_{\text{B}})
\rangle
=&\,\partial_{\tau_\text{A}}\partial_{\tau_\text{B}} g(\tau_\text{A}=0,
\tau_\text{B}=0) .
\end{align}
To derive higher-order correlation functions as in Eq. (\ref{eq:corr_general}),
$I_\text{A}$ and $I_\text{B}$ can be considered as a union of disjoint
subintervals $I_{\text{A}1}\ldots I_{\text{A}m_\text{A}}$ and $I_{\text{B}1}
\ldots I_{\text{B}m_\text{B}}$ containing the time values $T_{\text{A}1}\ldots 
T_{\text{A}m_\text{A}}$ and $T_{\text{B}1}\ldots T_{\text{B}m_\text{B}}$,
respectively. The subintervals designate independent modes, such that the
approximation in Eq. (\ref{eq:pointwise P_I}) can be generalized to
\begin{align}
P_{I_q} \approx \begin{pmatrix}
|I_{q1}\rangle\langle I_{q1}|&&0\\
&\ddots&\\
0&&|I_{q m_{q}}\rangle\langle I_{q m_{q}}|
\end{pmatrix}.
\end{align}
Reproducing the derivation of the PGF in Eq. (\ref{eq:g_small}) shows that it
suffices to make the substitutions
\begin{align}
\tau_q&\mapsto\begin{pmatrix}
\tau_{q1}&&0\\
&\ddots&\\
0&&\tau_{q m_{q}}
\end{pmatrix}
\end{align}
and
\begin{align}
\tilde{S}_{qq'}(T_q,T_{q'})&\mapsto\begin{pmatrix}
\tilde{S}_{qq'}(T_{q1},T_{q'1})&\cdots&\tilde{S}_{qq'}(T_{q1},T_{q'm_{q'}})\\
\vdots&\ddots&\vdots\\
\tilde{S}_{qq'}(T_{qm_{q}},T_{q'1})&\cdots&\tilde{S}_{qq'}
(T_{qm_{q}},T_{q'm_{q'}})
\end{pmatrix}
\end{align}
for each party $q,q'$ in Eq. (\ref{eq:g_small}). It can be seen that the
correlations between all time values $T_{\text{A}1}\ldots
T_{\text{A}m_\text{A}}$ and $T_{\text{B}1}\ldots T_{\text{B}m_\text{B}}$ are
taken into account. This exactly reveals a discretized version of the Fredholm
determinant in Eq. (\ref{eq:g_general}). By means of these substitutions,
general correlation functions in Eq. (\ref{eq:corr_general}) can be computed as
\begin{align}
\notag
\partial_{\tau_{\text{A}1}}\ldots\partial_{\tau_{\text{A}m_\text{A}}}
\partial_{\tau_{\text{B}1}}\ldots\partial_{\tau_{\text{B}m_\text{B}}}
g(&\tau_{\text{A}1}=0,\ldots,\tau_{\text{A}m_\text{A}}=0,
\\
&\tau_{\text{B}1}=0,\ldots,\tau_{\text{B}m_\text{B}}=0).
\end{align}

\subsection{Large interval widths}\label{s:spdc_large}
The intervals are now assumed to satisfy $|I_\text{A}|,|I_\text{B}|\gg\delta_t$.
The PGF in Eq. (\ref{eq:g_general}) can be written as
\begin{align}\label{eq:g_lndet}
g(y_\text{A},y_\text{B}) &= \exp\left\lbrace-\frac{1}{2}\text{Tr}\ln\left[
\mathbb{I} + P(y_\text{A},y_\text{B}) \left(\sigma_\text{f}-\frac{1}{2}
\mathbb{I} \right) \right]\right\rbrace
\end{align}
using the identity $\ln\det =\text{Tr}\ln $ \cite{Bornemann_2009}. According to
Eq. (\ref{eq:sigma_fiber}), the covariance $\sigma_\text{f}$ behaves
approximately diagonally compared to $I_\text{A}$ and $I_\text{B}$. Hence, it
commutes with the projection operators $P_{I_\text{A}}$ and $P_{I_\text{B}}$,
occurring in $P(y_\text{A},y_\text{B})$. Powers of $P(y_\text{A},y_\text{B}) 
\left(\sigma_\text{f}-\mathbb{I}/2 \right)$ in Eq. (\ref{eq:g_lndet}), however,
give rise to the combination $P_{I_\text{A}}P_{I_\text{B}} = P_{I_\text{A}\cap 
I_\text{B}}$ in some terms, where $I_\text{A}$ and $I_\text{B}$ might be
different. This problem is solved by the statement
\begin{widetext}
\begin{align}\label{eq:rule_P}
f\left[
\begin{pmatrix}
P_{I_\text{A}}&0\\0&P_{I_\text{B}}
\end{pmatrix}
\left(\sigma_\text{f}-\frac{1}{2}\mathbb{I} \right) \right]
\begin{pmatrix}
P_{I_\text{A}}&0\\0&P_{I_\text{B}}
\end{pmatrix}
\approx& \begin{pmatrix}
P_{I_\text{A}\setminus I_\text{B}} f\left[\left(\sigma_\text{f}-\mathbb{I}/2 
\right)_\text{AA}\right]
& 0 \\ 0 &
P_{I_\text{B}\setminus I_\text{A}} f\left[\left(\sigma_\text{f}-\mathbb{I}/2 
\right)_\text{BB}\right]
\end{pmatrix}
+
P_{I_\text{A}\cap I_\text{B}} f\left[\sigma_\text{f}-\frac{1}{2}\mathbb{I}
\right]
\end{align}
\end{widetext}
for analytic functions $f(x)$ fulfilling $f(0)=0$, which can be proven for the
power series of $f(x)$ by induction. Setting $f(x)\coloneqq \ln(1+x)$ in Eq.
(\ref{eq:rule_P}) leads to a decomposition
\begin{align}\label{eq:g_decomp}
g(y_\text{A},y_\text{B}) &=
g_\text{A}^{I_\text{A}\setminus I_\text{B}}(y_\text{A})\,
g_\text{B}^{I_\text{B}\setminus I_\text{A}}(y_\text{B})\,
g_{\text{j}}^{I_\text{A}\cap I_\text{B}}(y_\text{A},y_\text{B})
\end{align}
of the PGF in Eq. (\ref{eq:g_lndet}). The total counting statistics are thus
determined by the mutually uncorrelated statistics of Alice's counting at
$I_\text{A}\setminus I_\text{B}$, Bob's counting at $I_\text{B}\setminus
I_\text{A}$, and their correlation at $I_\text{A}\cap I_\text{B}$. The PGFs of
each party can be written as
$
g_\text{A}^{I}(y_\text{A}) = g_{\text{j}}^{I}(y_\text{A},1)
$ and $
g_\text{B}^{I}(y_\text{B}) = g_{\text{j}}^{I}(1,y_\text{B})
$
in terms of the joint PGF, given by
\begin{align}\label{eq:g_large_prev}
\notag
g_{\text{j}}^{I}(y_\text{A},y_\text{B})
= \exp &\left\lbrace -\frac{1}{2} \int_{I}\text{d}t\,\text{Tr}
\left[\left\lbrace
\ln \left[ \mathbb{I} +
\begin{pmatrix}
1-y_\text{A} & 0 \\ 0 & 1-y_\text{B}
\end{pmatrix}
\right.\right.\right.\right.\\&\left.\left.\left.\left.
\times\left(\sigma_\text{f}-\frac{1}{2}\mathbb{I}\right)\right]
\right\rbrace(t,t) 
\right] \right\rbrace.
\end{align}
In order to evaluate the operator at $(t,t)$, Eq. (\ref{eq:sigma_fiber}) can
be generalized to
\begin{align}\label{eq:fsigma_fS}
\notag
&\left[f\left(\sigma_\text{f}-\frac{1}{2}\mathbb{I}\right)\right](t,t')
\\&
= \frac{1}{\sqrt{2\pi}\delta_t} \mathcal{F}_x\left\lbrace f\left[ S\left(\frac{t
+t'}{2\Delta_t},x\right)\right] \right\rbrace\left(\frac{t-t'}{\delta_t}\right)
\end{align}
using theorem \ref{theorem}. It should be pointed out that fiber coupling might
have changed the degree of entanglement and thus the conditions of theorem
\ref{theorem}. Applying Eq. (\ref{eq:fsigma_fS}) to Eq. (\ref{eq:g_large_prev})
gives the desired result
\begin{align}\label{eq:g_large}
\notag
g_{\text{j}}^{I}(y_\text{A},y_\text{B})
=
\exp &\left\lbrace -\frac{\Delta_t}{4\pi\delta_t} \int_{I/\Delta_t}\text{d}\chi
\,\int_{-\infty}^\infty\text{d}\kappa
\right.\\&\left.
\ln \det\left[ \mathbb{I} +
\begin{pmatrix}
1-y_\text{A} & 0 \\ 0 & 1-y_\text{B}
\end{pmatrix}
S(\chi,\kappa) \right] \right\rbrace,
\end{align}
where $S(\chi,\kappa)$ is determined by Eq. (\ref{eq:sigma_fiber}). In contrast
to Eq. (\ref{eq:g_small}), the integrals are outside of the determinant and the
logarithm. Some insights in this observation may be gained by extracting the
integrals from the exponential function, which would reveal a product integral
(a continuous version of a product) of the determinant. Hence, the correlations
of all combinations of parties and values for $\chi,\kappa$ are taken into
account. As in Sec. \ref{s:spdc_small}, this formula is efficiently computable
due to limited size of $S(\chi,\kappa)$, even for increasing entanglement.
Moreover, Eq. (\ref{eq:g_large}) again serves to generalize the counting
statistics to further components and external influences, as discussed in Sec.
\ref{s:General Procedure}. If the phase matching is collinear degenerate and the
entangled photon pairs are detected without further influences, the PGF in Eq.
(\ref{eq:g_large}) simplifies to
\begin{align}\label{eq:g_large_direct}
g_{\text{j}}^{I}(y_\text{A},y_\text{B})
=& \exp\left\lbrace -\frac{\Delta_t}{2\pi\delta_t} \int_{I/\Delta_t}\text{d}\chi
\,\int_{-\infty}^\infty\text{d}\kappa
\right.\\\notag&\qquad\left.
\ln \left[ 1 +\frac{1-y_\text{A}y_\text{B}}{2} \phi_\text{c}(\chi,\kappa) 
\right] \right\rbrace,
\end{align}
where $\phi_\text{c}(\chi,\kappa)$ is defined in Eq. (\ref{eq:phi_c,s}). It can
be inferred that the statistics are strictly correlated since the PGF only
depends on the product $y_\text{A}y_\text{B}$, which matches with
\cite{Mauerer_2009,Burenkov_2017}.

For one party (e.g., Alice) and collinear degenerate phase matching, the
statistics become Poissonian in the case of perfect entanglement. An expansion
about this case gives
\begin{align}\label{eq:g_A high}
g_\text{A}(y) \approx \exp\left[(y-1)\langle\hat{N}\rangle + \frac{(y-1)^2}{2}
\frac{\langle\hat{N}\rangle^2}{K}\right]
\end{align}
revealing the Schmidt number $K$ and the mean photon number $\langle\hat{N}
\rangle$, determined by Eqs. (\ref{eq:K}) and (\ref{eq:N}), respectively. This
succinctly describes how far the degree of entanglement influences the counting
statistics of one party.

\subsection{Intermediate interval widths}
The remaining case, where the approximations applied to the projection operators
$P_{I_\text{A}},P_{I_\text{B}}$ in the latter sections are not sufficient
anymore, turns out to be more difficult. Since an analytical solution without
any approximations is hardly possible, intermediate interval widths can only be
investigated by expanding the limiting cases in terms of interval widths,
benefiting a more detailed description of the counting statistics. In
particular, large interval widths, modeled by $\sigma_\text{f}$ being
approximately diagonal compared to $P_{I_\text{A}},P_{I_\text{B}}$ and discussed
in Sec. \ref{s:spdc_large}, immediately lead to the fact that correlations
solely exist between overlapping intervals, as presented in Eq.
(\ref{eq:g_decomp}). Due to finite temporal correlation widths $\delta_t$,
however, correlated photons are able to overcome the interval borders and to be
detected at separated time intervals, most significantly if the intervals are
neighbored. Hence, counting statistics that cover this influence are of
considerable interest for experimental applications. It is envisioned to extend
the PGF in Eq. (\ref{eq:g_decomp}) by expansion about the approximation for
large interval widths and to gain potential correlations between separated time
intervals. It should be mentioned that this expansion is inaccessible to
approaches based on Schmidt decomposition or general optical modes since
diagonal $\sigma_\text{f}$ corresponds to perfect entanglement.

In view of Eq. (\ref{eq:g_lndet}), it is worthwhile to express the logarithm in
its power series and to consider each power $n\in\mathbb{N}$ as
\begin{widetext}
\begin{align}\label{eq:p_sigma_n}
\text{Tr}\left\lbrace\left[P(y_\text{A},y_\text{B}) \left(\sigma_\text{f}-
\frac{1}{2}\mathbb{I} \right)\right]^n\right\rbrace
= \sum_{
\substack{q_1,\ldots q_n \\ =\text{A,B}}
} \int\text{d}t_1\ldots\text{d}t_n & \text{Tr}\left\lbrace (1-y_{q_1})
\textbf{1}_{t_1\in I_{q_1}}\left[\sigma_\text{f}-\frac{1}{2}\mathbb{I} 
\right]_{q_1,q_2}(t_1,t_2)
\right.\notag\\&\ldots\left.
(1-y_{q_n})\textbf{1}_{t_n\in I_{q_n}}\left[\sigma_\text{f}-\frac{1}{2}
\mathbb{I} \right]_{q_n,q_1}(t_n,t_1)\right\rbrace.
\end{align}
\end{widetext}
To derive Eq. (\ref{eq:rule_P}), each projection operator was commuted and
combined with the first one, which corresponds to the approximation
$\textbf{1}_{t_j\in I_{q_j}}\approx\textbf{1}_{t_1\in I_{q_j}}$ for all
$j=2\ldots n$. For a more detailed description, the error is taken into account
as
\begin{align}
\textbf{1}_{t_j\in I_{q_j}}=\textbf{1}_{t_1\in I_{q_j}}+\varepsilon_{q_j}(t_j).
\end{align}
Expanding the product of indicator functions in Eq. (\ref{eq:p_sigma_n}) w.r.t.
$\varepsilon_{q_j}(t_j)$ as
\begin{align}
\prod_{j=1}^n \textbf{1}_{t_j\in I_{q_j}} &\approx \textbf{1}_{t_1\in 
\bigcap_{j=1}^nI_{q_j}} + \sum_{k=2}^n \textbf{1}_{t_1\in \bigcap_{j\neq k}
I_{q_j}}\varepsilon_{q_k}(t_j)
\end{align}
gives the first order of the desired case of intermediate interval widths. It is
worth noting that each interval $I_{q_j}$ corresponds to either $I_\text{A}$ or
$I_\text{B}$. It is thus pertinent to distinguish whether all intervals
coincide, only one interval differs, or at least two intervals differ. Depending
on these cases, $\textbf{1}_{t_1\in\bigcap_{j=1}^n I_{q_j}}$ and
$\textbf{1}_{t_1\in\bigcap_{j\neq k}I_{q_j}}$ equal to either
$\textbf{1}_{t_1\in I_\text{A}\cap I_\text{B}}$ or
$\textbf{1}_{t_1\in I_{q_1}}$.

A detailed calculation using Eq.
(\ref{eq:sigma_fiber}) reveals an extended version of Eq. (\ref{eq:rule_P}),
which leads to the sought-after PGF
\begin{align}
g(y_\text{A},y_\text{B}) =&
g_{\text{j}}^{I_\text{A}\cap I_\text{B}}(y_\text{A},y_\text{B})\,
g_\text{A}^{I_\text{A}\setminus I_\text{B}}(y_\text{A})\,
g_\text{B}^{I_\text{B}\setminus I_\text{A}}(y_\text{B})
\notag\\&
\times g_{\text{A,cor}}^{I_\text{A}\setminus I_\text{B}}(y_\text{A},y_\text{B})
\,
g_{\text{B,cor}}^{I_\text{B}\setminus I_\text{A}}(y_\text{A},y_\text{B}).
\end{align}
It can be seen that the functions $g_{\text{A,cor}}$ and $g_{\text{B,cor}}$
provide correlations between separated intervals. Once again, the PGFs of each
party can be written as
$
g_\text{A}^{I}(y_\text{A}) = g_{\text{j}}^{I}(y_\text{A},1)
$ and $
g_\text{B}^{I}(y_\text{B}) = g_{\text{j}}^{I}(1,y_\text{B})
$
in terms of the joint PGF
\begin{widetext}
\begin{align}\label{eq:g_inter_joint}
g_{\text{j}}^{I}(y_\text{A},y_\text{B}) &=
\exp \left( -\frac{\Delta_t}{4\pi\delta_t} \int_{I/\Delta_t}\text{d}\chi\,
\int_{-\infty}^\infty\text{d}\kappa\,
\text{Tr}\left\lbrace\ln \left[ \mathbb{I} +
\begin{pmatrix}
1-y_\text{A} & 0 \\ 0 & 1-y_\text{B}
\end{pmatrix}
S(\chi,\kappa) \right]
\right.\right.\notag\\&\left.\left.
+ \sum_q P_q \mathcal{F}_{\kappa'}\left\lbrace
\varepsilon_q(\kappa')  f\left[\begin{pmatrix}
1-y_\text{A} & 0 \\ 0 & 1-y_\text{B}
\end{pmatrix} S\left(\chi,\kappa+\kappa'\right),\begin{pmatrix}
1-y_\text{A} & 0 \\ 0 & 1-y_\text{B}
\end{pmatrix} S\left(\chi,\kappa-\kappa'\right)\right]
\right\rbrace\left(2\frac{\Delta_t\chi-\overline{T}_q}{\delta_t}\right)
\right\rbrace\right),
\end{align}
\end{widetext}
where $P_q$ and $\overline{T}_q$ denote a $2\times 2$ projection matrix onto
party $q$ and the midpoint of $I_q$, respectively, and $S(\chi,\kappa)$ is
determined by Eq. (\ref{eq:sigma_fiber}). The error is represented by its
Fourier transformation
\begin{align}\label{eq:epsilon_kappa'}
\varepsilon_q\left(\kappa'\right)\coloneqq \sqrt{2\pi}\left[ \frac{|I_{q}|}{\pi
\delta_t}\,\text{sinc}\left(\kappa'\frac{|I_{q}|}{\delta_t}\right) - 
\delta(\kappa')\right],
\end{align}
which vanishes for $|I_{q_k}|/\delta_t \rightarrow\infty$. Moreover, the matrix-
valued function
\begin{align}
f(M,N)\coloneqq \sum_{n=1}^{\infty} \frac{(-1)^{n+1}}{n}\sum_{k=1}^{n-1} M^{k} 
N^{n-k}
\end{align}
for matrices $M,N$, originating from the power series of the logarithm, does not
generally simplify to an analytic expression since $S\left(\chi,\kappa+
\kappa'\right)$ and $S\left(\chi,\kappa-\kappa'\right)$ do not commute. This is
the case for one party though, such that
\begin{align}\label{eq:f(M,N)_analytic}
f(M,N) = (M-N)^{-1}[N\ln(\mathbb{I}+M)-M\ln(\mathbb{I}+N)]
\end{align}
can be applied to compute $g_\text{A}^{I}(y_\text{A})$ and $g_\text{B}^{I}
(y_\text{B})$. It should be noted that the first term in Eq.
(\ref{eq:g_inter_joint}) corresponds to the joint PGF
for large interval widths in Eq. (\ref{eq:g_large}). The correction PGF is
given by
\begin{widetext}
\begin{align}\label{eq:g_inter_cor}
g_{q\text{,cor}}^{I}(y_\text{A},y_\text{B}) = \exp&\left\lbrace
-\frac{\Delta_t}{4\pi\delta_t} \int_{I/\Delta_t}\text{d}\chi\,
\int_{-\infty}^\infty\text{d}\kappa\,\text{Tr}\,\mathcal{F}_{\kappa'}
\left\lbrace
\varepsilon_{\neg q}\left(\kappa'\right) (1-y_q)S_{q\neg q}(\chi,\kappa+\kappa') 
(1-y_{\neg q})S_{\neg qq}(\chi,\kappa-\kappa')
\right.\right.\notag\\&\left.\left.
\times\tilde{f}\left[(1-y_q)S_{qq}(\chi,\kappa-\kappa'),(1-y_q)S_{qq}(\chi,
\kappa+\kappa')\right]
\right\rbrace \left(2\frac{\Delta_t\chi-\overline{T}_q}{\delta_t}\right)
\right\rbrace,
\end{align}
\end{widetext}
where $\neg q$ denotes the opposite party of $q$ and $\tilde{f}(M,N)\coloneqq 
M^{-1}f(M,N)N^{-1}$ can be expressed using Eq. (\ref{eq:f(M,N)_analytic}) since
$S_{qq}(\chi,\kappa-\kappa')$ and $S_{qq}(\chi,\kappa+\kappa')$ commute for each
party $q$.

The presented results can be applied to coding in time basis, where the pump
pulse consists of two pulses that are localized in temporally separated
intervals. The establishment of the key bits is based on the perturbative limit,
where exactly one photon pair is assumed to be generated and the probability of
Alice and Bob detecting these photons at different intervals would thus vanish.
A more detailed description is already given by the PGF of large interval widths
in Eq. (\ref{eq:g_decomp}), which, however, only covers uncorrelated photon
pairs. Correlations between these intervals due to finite $\delta_t$ are
provided in Eqs. (\ref{eq:g_inter_joint}) and (\ref{eq:g_inter_cor}).
To evaluate these latter PGFs, the probability of each party detecting exactly
one photon at separated time intervals $I_\text{A},I_\text{B}$ is addressed,
which is the central figure of merit for coding in time basis and is given by
\begin{align}
p(n_\text{A}=1,I_\text{A};n_\text{B}=1,I_\text{B})
=& \partial_\text{A} g_\text{A}^{I_\text{A}}(0)\,\partial_\text{B}
g_\text{B}^{I_\text{B}}(0)
\notag\\&+ \partial_\text{A}\partial_\text{B}g_{\text{A,cor}}^{I_\text{A}}(0,0)
\notag\\&+ \partial_\text{A}\partial_\text{B}g_{\text{B,cor}}^{I_\text{B}}(0,0).
\end{align}
It can be noted that the first term represents the detection of uncorrelated
photon pairs. For the sake of simplicity, the intervals are set to be
$I_\text{A}\coloneqq[0,\infty)$ and $I_\text{B}\coloneqq(-\infty,0]$, which
yields the corresponding PGFs as
\begin{widetext}
\begin{align}\label{eq:g_inter_appl}
g_{q}^{I_q}(y) = \exp&\left(
-\frac{\Delta_t}{4\pi\delta_t}\int_{I_q/\Delta_t}\text{d}\chi\,
\int_{-\infty}^\infty\text{d}\kappa \left\lbrace 2\ln\left[ 1+\frac{1-y}{2}
\phi_\text{c}(\chi,\kappa) \right] 
- \int_{-\infty}^\infty \text{d}\kappa'\, |\chi|\,\tilde{\varepsilon}
\left(|\chi|\kappa'\right)
\right.\right.\notag\\&\quad\left.\left.
\times\frac{
\ln\left[1+\frac{1-y}{2}\phi_\text{c}(\chi,\kappa+\kappa')\right]\phi_\text{c}
(\chi,\kappa-\kappa')-\ln\left[1+\frac{1-y}{2}\phi_\text{c}(\chi,\kappa-\kappa')
\right]\phi_\text{c}(\chi,\kappa+\kappa')
}{\phi_\text{c}(\chi,\kappa+\kappa')-\phi_\text{c}(\chi,\kappa-\kappa')}
\right\rbrace\right),
\notag\\[0.5em]
g_{q\text{,cor}}^{I_q}(y_\text{A},y_\text{B}) = \exp&\left\lbrace
-\frac{\Delta_t}{4\pi\delta_t}\frac{1-y_{\neg q}}{2} \int_{I_q/\Delta_t}\text{d}
\chi\,\int_{-\infty}^\infty\text{d}\kappa\,\text{d}\kappa'\, |\chi|\,
\tilde{\varepsilon}\left(|\chi|\kappa'\right)
\right.\notag\\&\quad\left.
\times\Re\left[\phi_\text{s}(\chi,\kappa+\kappa')\phi_\text{s}(\chi,\kappa-
\kappa')^*\right]\frac{
f\left[\frac{1-y_q}{2}\phi_\text{c}(\chi,\kappa+\kappa')\right]-f\left[\frac{1-
y_q}{2}\phi_\text{c}(\chi,\kappa-\kappa')\right]
}{\phi_\text{c}(\chi,\kappa+\kappa')-\phi_\text{c}(\chi,\kappa-\kappa')}
\right\rbrace,
\end{align}
where $f(x)\coloneqq \ln(1+x)/x$ and $\phi_\text{c}(\chi,\kappa),
\phi_\text{s}(\chi,\kappa)$ are defined in Eq. (\ref{eq:phi_c,s}).
\end{widetext}
The error is now given by
\begin{align}
\tilde{\varepsilon}(x)\coloneqq \delta(x)-\frac{2\Delta_t}{\pi\delta_t}\,
\text{sinc}\left(\frac{2\Delta_t}{\delta_t}x\right),
\end{align}
where, compared to Eq. (\ref{eq:epsilon_kappa'}), the pulse width plays the role
of the interval width, which is attributed to the fact that detection outside
the pulse width is negligible. The error thus includes the degree of
entanglement and vanishes for $\delta_t/\Delta_t\rightarrow 0$. Conversely, the
term $|\chi|\,\tilde{\varepsilon}\left(|\chi|\kappa'\right)$, occurring in Eq.
(\ref{eq:g_inter_appl}), depends on the distance to the interval border
$|\chi|$, which the photon pairs need to overcome in order to trigger the
detection of correlations. Hence, the error is maximal for $\chi=0$ and vanishes
for $|\chi|\rightarrow\infty$.

\begin{figure}
\centering
\includegraphics[width=0.45\textwidth]{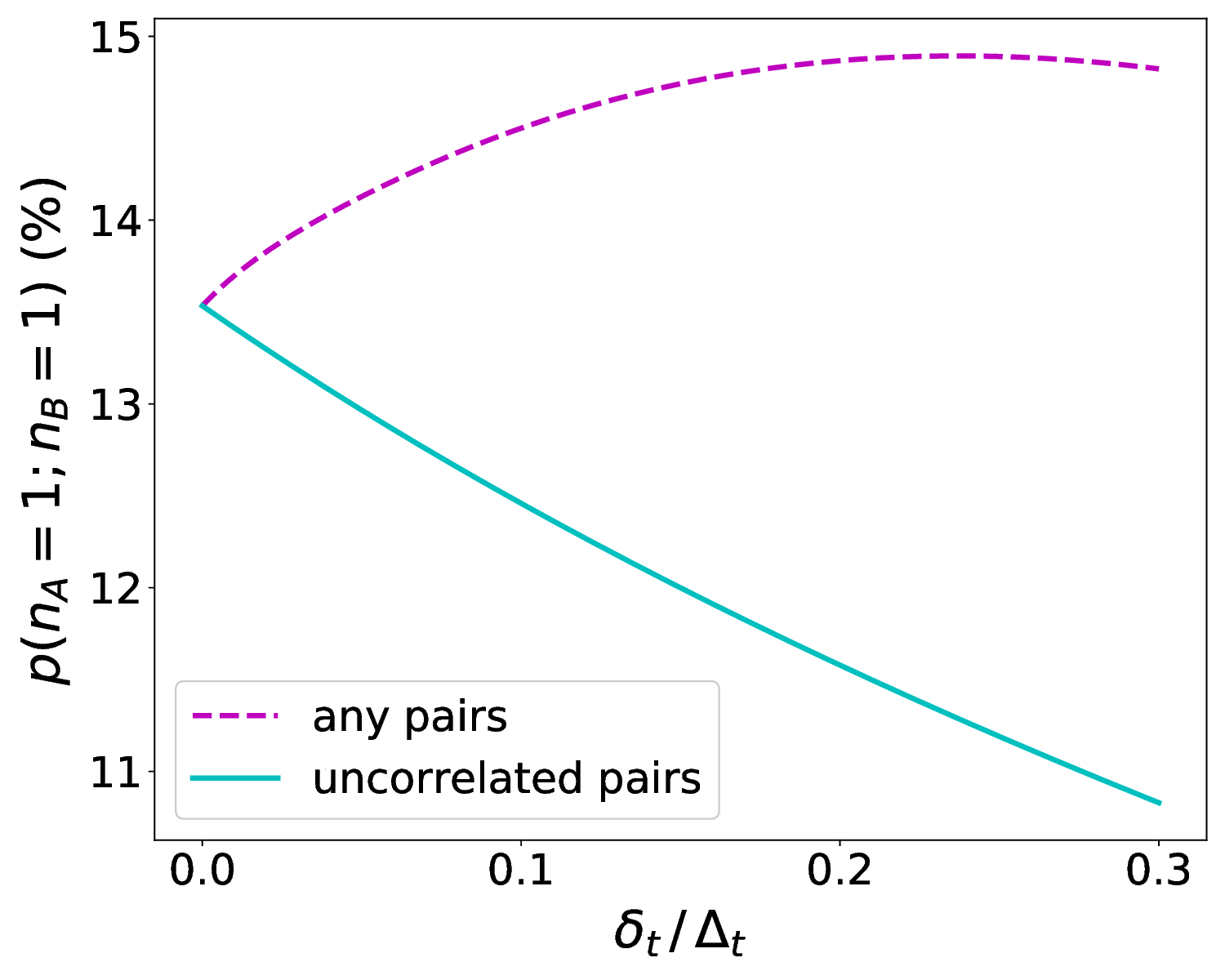}
\caption{Probability of each party detecting exactly one photon in dependence on
the temporal correlation width $\delta_t$. The cases of uncorrelated photon
pairs and any (correlated and uncorrelated) photon pairs are displayed. The
probability to detect correlated photon pairs vanishes at $\delta_t=0$,
increases for higher values of $\delta_t/\Delta_t$, and starts to decrease again
for $\delta_t/\Delta_t\gtrsim 0.3$. Since each party's statistics depend on the
degree of entanglement, the probability to detect uncorrelated photons
decreases.
}
\label{fig:disjoint_intervals}
\end{figure}

The probability of each party detecting exactly one photon in dependence on the
temporal correlation width $\delta_t$ is illustrated in Fig.
\ref{fig:disjoint_intervals}, where the detection of uncorrelated and any
(uncorrelated and correlated) photon pairs is displayed. For the sake of
simplicity, the mean photon number of the entire pulse is set to be
$\langle\hat{N}\rangle=2$, such that each interval $I_q$ contains
$\langle\hat{N}_{q}\rangle=1$. What is evident from this figure is that the
probability for the detection of correlated photon pairs, corresponding to the
difference of the curves, vanishes at $\delta_t=0$ and increases for higher
values of $\delta_t/\Delta_t$. It can be inferred that correlated photon pairs
have a modest, yet not negligible, influence on the detection results. The
decrease of the probability to detect uncorrelated photons can straightforwardly
be understood based on the fact that the statistics depend on the degree of
entanglement, being Poissonian for perfect entanglement and thermal without
entanglement. For $\delta_t/\Delta_t\gtrsim 0.3$, the curve depicting the
detection of any photon pairs starts to decrease, which could suggest that
expanding the limiting case of large interval widths within the first order is
not sufficient anymore and further orders need to be taken into account, which
could lead to a further increase. A detailed computation, however, would be
beyond the scope of this paper and might be addressed elsewhere. Moreover, it
should be kept in mind that the approach of this paper is only valid for a
certain range of entanglement, as presented in Tab. \ref{tab:N_max}.

\section{Setups and external influences}\label{s:setup,influences}
So far, the entangled photon pairs were considered to be detected without any
setup or environmental influences. In this section, the results are generalized
to further optical components and external influences of interest for
experimental realizations. To this end, a general procedure is presented and
applied to phase-time coding.

\subsection{General procedure}\label{s:General Procedure}
Linear optical processes can be modeled by a unitary operator $F$ modifying
the covariance $\sigma\mapsto F\sigma F^\text{T}$ \cite{Olivares_2012}. Based
on Eqs. (\ref{eq:sigma_S}) and (\ref{eq:sigma_fiber}), this corresponds to a
transformation of the matrices $S(\boldsymbol{\chi},\boldsymbol{\kappa})$ and
$S(\chi,\kappa)$ for free space and fiber propagation, respectively. In the
latter case, the PGFs for small, large, and intermediate interval widths,
given by Eqs. (\ref{eq:g_small}), (\ref{eq:g_large}), (\ref{eq:g_inter_joint}),
and (\ref{eq:g_inter_cor}), respectively, can simply be generalized by modifying
$S(\chi,\kappa)$. In the case of a modular array of several
components and influences, the corresponding transformations can be applied
successively.

There is only need to clarify whether the corresponding assumptions are still
verified. In general, the condition of theorem \ref{theorem} needs to be checked
again and, if it is compromised, Tab. \ref{tab:N_max} needs to be adjusted.
Moreover, in the case of small interval widths, derived in Sec.
\ref{s:spdc_small},
\begin{enumerate}
\item\label{cond:1} $F\sigma F^\text{T}$ still needs to be approximately
constant compared to $I_\text{A},I_\text{B}$.
\end{enumerate}
\indent
In the case of large and intermediate interval widths, derived in Sec.
\ref{s:spdc_large},
\begin{enumerate}
\item $F\sigma F^\text{T}$ still needs to behave approximately diagonally
compared to $I_\text{A},I_\text{B}$ and
\item\label{cond:2} the transformation of $S(\chi,\kappa)$, derived based on Eq.
(\ref{eq:sigma_fiber}), has to be extended to the generalized version in Eq.
(\ref{eq:fsigma_fS}).
\end{enumerate}
Conditions \ref{cond:1} can simply be verified by determining the maximal
variation and the shape of $F\sigma F^\text{T}$, respectively. To verify
condition \ref{cond:2}, it is suggested to investigate, whether $F$ completely
or partially commutes with $\sigma$.

This procedure can be applied to various setups and external influences. Apart
from phase-time coding, which is discussed in the next section, full
descriptions of Hong-Ou-Mandel interferometric measurement, nonlinear
dispersion, and polarization mode dispersion were derived. A detailed
presentation, however, would be beyond the scope of this paper and will be
addressed elsewhere.

\subsection{Phase-time coding}\label{s:Phase-time coding}
In this section, the PGF for large interval widths, given by Eq.
(\ref{eq:g_large}), is applied to the setup of phase-time coding with fiber
propagation and the detuning of the MZIs affecting Franson interference is
investigated.

The pump pulse consists of two pulses $\alpha_+$ and $\alpha_-$ with a temporal
delay $\tau$ and a phase difference $\varphi_\alpha$. The setup comprises a MZI
at each party's side with time delays $\tau_\text{A},\tau_\text{B}$ and phase
shifts $\varphi_\text{A},\varphi_\text{B}$, respectively. The corresponding
detection profile consists of three pulses that are localized in time intervals
$I_\text{s}$, $I_\text{m}$, and $I_\text{l}$. The interval $I_\text{s}$
($I_\text{l}$) is attributed to the early (late) pump pulse taking the short
(long) arm of the MZI. The middle interval $I_\text{m}$ contains a superposition
of the early pump pulse propagating through the long arm and of the late pump
pulse propagating through the short arm. Phase-time coding is based on two
assumptions \cite{Brendel_1998}: the generation of exactly one photon pair and
perfect Franson interference between $I_\text{m}$ and $I_\text{m}$. The first
statement is merely a rough estimation, which is apparent from Sec.
\ref{s:spdc_large}. The second statement, where the correlation is assumed to
vanish in the case of $\varphi=\pi$ with $\varphi\coloneqq\varphi_\alpha-
\varphi_\text{A}-\varphi_\text{B}$, is closely related to the first one. It
turns out that destructive interference $\varphi=\pi$ and constructive
interference $\varphi=0$ lead to uncorrelated and strictly correlated counting
statistics, respectively, which are both alleviated by attenuation. To this end,
multiple photon pairs need to be taken into account, which excludes the
perturbative limit to describe Franson interference suitably. Moreover, the
second condition holds only if the time delays of the MZIs coincide, which,
however, is already affected if they differ in the magnitude of the temporal
correlation width $\delta_t$ \cite{Gao_2019}. As it becomes apparent shortly,
the dependence on the phase differences $\varphi$ is already negligible for a
detuning $|\tau_\text{A}-\tau_\text{B}|\geq\delta_t$. Since the security of the
key relies on perfect Franson interference, a detailed description of this
influence is indispensable.

To investigate the full contribution of Franson interference, $\alpha_-=\alpha_+ 
e^{i\varphi_\alpha}$ and collinear degenerate phase matching are assumed.
Moreover, the phase matching is considered to be symmetric, that is, $\phi(\chi,
\kappa)$, defined in Eq. (\ref{eq:ja_t}), is symmetric w.r.t. $\kappa$. Only one
choice of detectors for each party $q$ is investigated since the opposite
detector can be modeled by shifting the phase $\varphi_q\mapsto\varphi_q+\pi$.
The pulses $\alpha_+$, $\alpha_-$ and the three pulses of the detection profile
are assumed to be completely separated, such that $+,-$ and $I_\text{s},
I_\text{m},I_\text{l}$ designate independent modes. In particular, this means
\begin{align}\label{eq:tau_cond}
\vert\tau_\text{A}-\tau\vert,\,
\vert\tau_\text{B}-\tau\vert,\,
\vert\tau_\text{A}-\tau_\text{B}\vert\ll\Delta_t .
\end{align}
The covariance of the biphoton state is thus given by
\begin{align}
\sigma = \begin{pmatrix}
\sigma_+ & 0 \\ 0 & \sigma_-
\end{pmatrix} ,
\end{align}
where $\sigma_\pm$ depends on the pump amplitude $\alpha_\pm$. The MZI of party
$q$ can be modeled by
\begin{align}
F_q = \sum_\pm \sqrt{t_{q\pm}}\mathcal{R}_{\pm{\varphi_q}/{2}}U_{\pm{\tau_q}/
{2}},
\end{align}
where $\mathcal{R}$ and $t_{q+},t_{q-}$ denote a $2\times 2$ rotation matrix
with subscripted rotation angle and the transmissions of each arm including the
BSs eliminating one outcome each, respectively. The time shift operators
$U_{\pm{\tau_q}/{2}}$ can be written as a transformation of the modes $+,-$ to
$I_\text{s},I_\text{m},I_\text{l}$:
\begin{align}
U_{\tau_q/{2}} &= \begin{pmatrix}
1 & 0 \\ 0 & 1 \\ 0 & 0
\end{pmatrix}
U_{(\tau_q-\tau)/{2}}
,\;
U_{-\tau_q/{2}} = \begin{pmatrix}
0 & 0 \\ 1 & 0 \\ 0 & 1
\end{pmatrix}
U_{-(\tau_q-\tau)/{2}} ,
\end{align}
where the residual time shifts $U_{\pm(\tau_q-\tau)/2}$ represent the delay
difference between the MZI of party $q$ and the pump pulses. To apply $F_q$ to
the PGF for large interval widths, given by Eq. (\ref{eq:g_large}), the
corresponding transformation of $S(\chi,\kappa)$ can be derived based on Eq.
(\ref{eq:sigma_fiber}). Since the finite matrix structure of $F_q$ remains
unchanged, only the residual time shifts need to be addressed. It can be shown
that a time shift operator $U_T$ modifying $\sigma_{qq'}\mapsto U_T\sigma_{qq'}$
or $\sigma_{qq'}\mapsto \sigma_{qq'} U_T$ for each party $q,q'$ can be
substituted by a phase shift
\begin{align}\label{eq:S(x',x) time_shift}
S_{qq'}(\chi,\kappa)\mapsto e^{i\kappa T/\delta_t}S_{qq'}(\chi,\kappa)
\end{align}
if $T\ll\Delta_t$ is fulfilled. Owing to this condition, $\chi$ does not need to
be shifted. In the case of the residual time shifts $U_{\pm(\tau_q-\tau)/2}$,
this condition is verified due to Eq. (\ref{eq:tau_cond}). The conditions to
apply this transformation, presented in Sec. \ref{s:General Procedure}, are also
satisfied owing to $\vert \tau_q-\tau\vert\ll\vert I_\text{m}\vert$ for each
party $q$ and since $U_{\pm(\tau_q-\tau)/2}$ and $\sigma$ approximately commute.
Applying Eq. (\ref{eq:S(x',x) time_shift}) and setting $I_\text{A}=I_\text{B}=
I_\text{m}$ gives
\begin{align}
\notag
S(\chi,\kappa)&\mapsto \sum_\pm
F_\pm(\kappa) S_\pm(\chi,\kappa)F_\pm(\kappa)^\dagger,
\\\notag
F_\pm(\kappa)&\coloneqq \begin{pmatrix}
F_{\text{A}\pm}(\kappa)&0\\0&F_{\text{B}\pm}(\kappa)
\end{pmatrix},
\\
F_{q\pm}(\kappa)&\coloneqq
\sqrt{t_{q\mp}} e^{\mp i\kappa(\tau_q-\tau_{\neg q})/(4\delta_t)}
\mathcal{R}_{\mp{\varphi_{q}}/{2}},
\end{align}
where $\neg q$ denotes the opposite party of $q$ and $S_\pm(\chi,\kappa)$
depends on the pulse $\alpha_\pm$. Setting $t_{q\pm}=1/4$, which is determined
by tracing out an output port of each BS, facilitates the PGF to give
\begin{widetext}
\begin{align}\label{eq:g_coding}
\notag
g(y_\text{A},y_\text{B})
=&\exp\left[-\frac{\Delta_t}{4\pi\delta_t} \int_{I_\text{m}/\Delta_t}\text{d}
\chi\,\int_{-\infty}^\infty\text{d}\kappa
\ln \left(\left\lbrace 1+\frac{1-y_\text{A}+1-y_\text{B}}{4}\phi_\text{c}(\chi,
\kappa)
+ \frac{1-y_\text{A}}{4}\frac{1-y_\text{B}}{4}\phi_\text{c}(\chi,\kappa)
\right.\right.\right.\\\notag
&\times\left.\left.\left.
\left[ \frac{\phi_\text{c}(\chi,\kappa)}{2}-1-\left( \frac{\phi_\text{c}(\chi,
\kappa)}{2}+1
\right) \cos\varphi\cos\frac{\tau_\text{A}-\tau_\text{B}}{\delta_t}\kappa\right] 
\right\rbrace^2
\right.\right.\\
&
-\left.\left.\left\lbrace \frac{1-y_\text{A}}{4}\frac{1-y_\text{B}}{4} 
\frac{\phi_\text{c}(\chi,\kappa)+2}{2}\phi_\text{c}(\chi,\kappa)
\sin\varphi\sin\frac{\tau_\text{A}-\tau_\text{B}}{\delta_t}\kappa
\right\rbrace^2 \right)\right],
\end{align}
\end{widetext}
where $\phi_\text{c}(\chi,\kappa)$, defined in Eq. (\ref{eq:phi_c,s}), is
symmetric w.r.t. $\kappa$ due to symmetric phase matching. It should be noted
that the results do not depend on $\tau$ but only on $\tau_\text{A}-
\tau_\text{B}$, even though $\tau$ is still limited due to Eq.
(\ref{eq:tau_cond}). In the case of no detuning $\tau_\text{A}=\tau_\text{B}$,
Franson interference becomes the most visible. Constructive interference
$\varphi=0$ yields strictly correlated counting statistics of Alice and Bob
including attenuation $t_\text{A}=t_\text{B}=1/2$, given by Eqs.
(\ref{eq:g_large}) and (\ref{eq:g_ineff}). Destructive interference
$\varphi=\pi$ leads to uncorrelated counting statistics, where the statistics of
each party coincide to the case of constructive interference.

\begin{figure}
\centering
\includegraphics[width=0.45\textwidth]{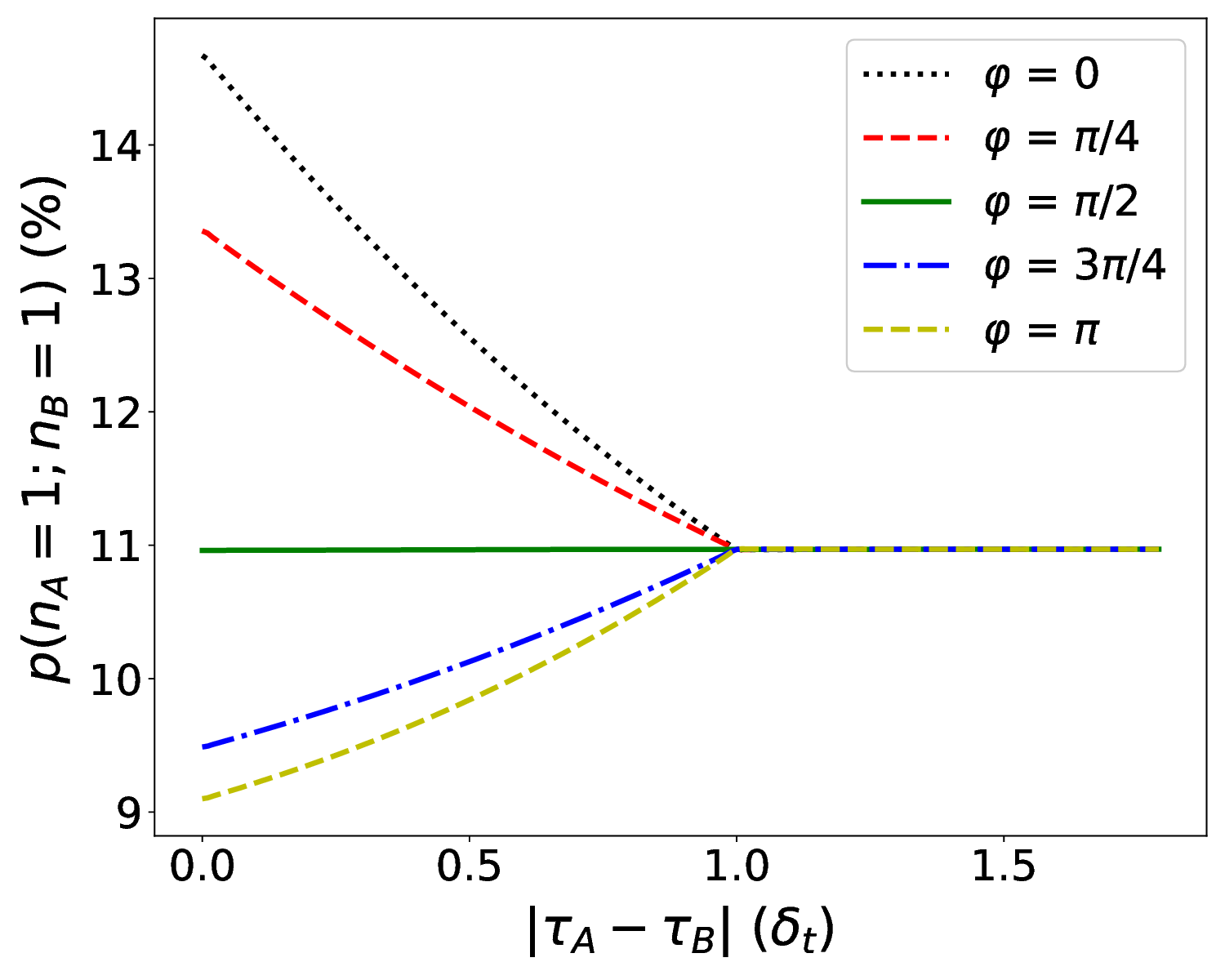}
\caption{Probability of Alice and Bob detecting both exactly one photon at the
intervals $I_\text{A}=I_\text{B}=I_\text{m}$ in dependence on the detuning
$\tau_\text{A}-\tau_\text{B}$ and the phase difference $\varphi$. Franson
interference becomes the most visible in the case of no detuning. For increasing
detuning, the probabilities converge to the constant curve of $\varphi=\pi/2$
and reach the limit at $\delta_t$. The dependence on the detuning is mainly
linear and yet asymmetric compared to the constant limit curve, caused by a
modest curvature.
}
\label{fig:detuning_exact1}
\end{figure}

\begin{figure}
\centering
\includegraphics[width=0.45\textwidth]{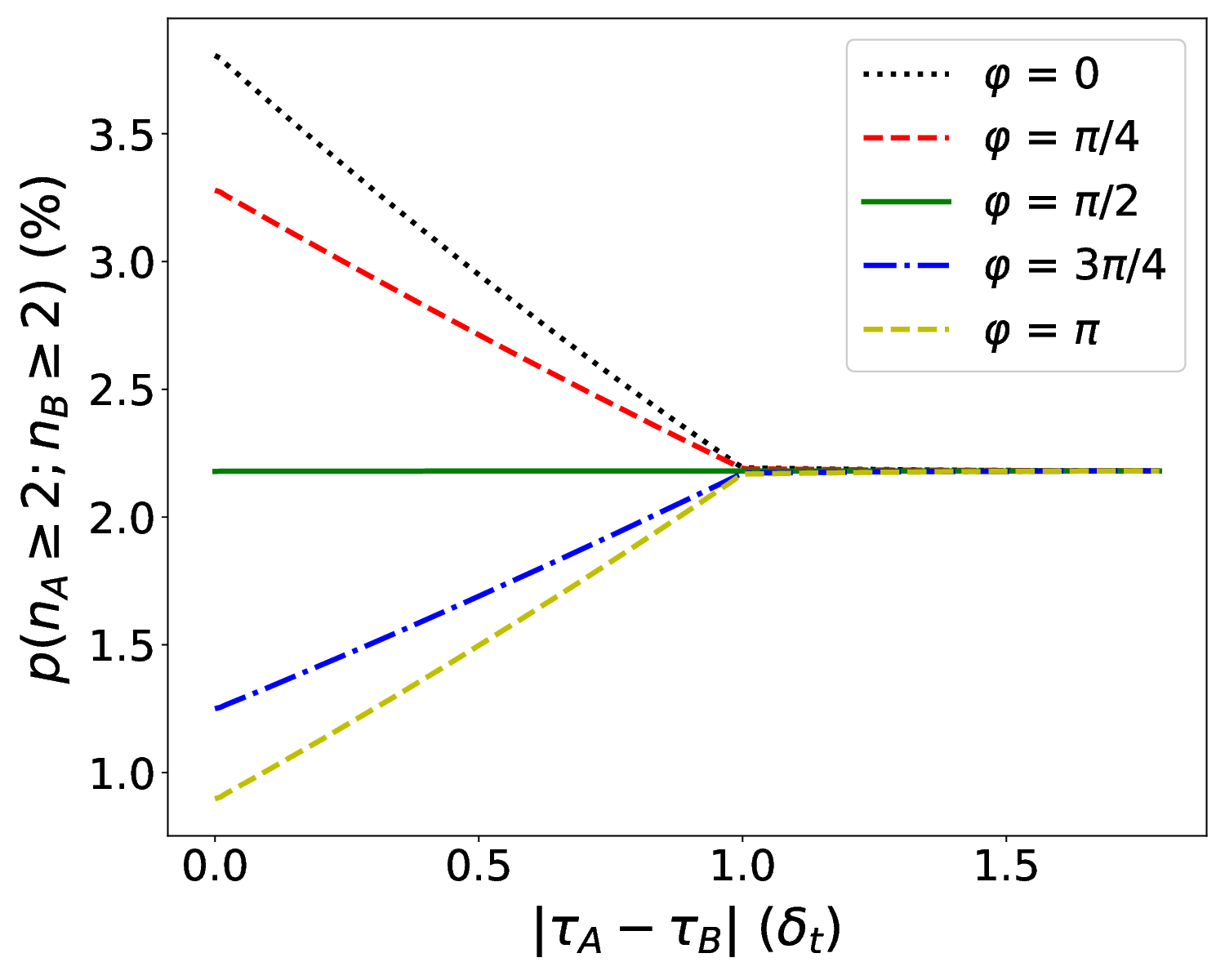}
\caption{Probability of Alice and Bob detecting both more than one photon at the
intervals $I_\text{A}=I_\text{B}=I_\text{m}$ in dependence on the detuning
$\tau_\text{A}-\tau_\text{B}$ and the phase difference $\varphi$. Similar curve
shapes arise as in Fig. \ref{fig:detuning_exact1} at lower probabilities and
with lower curvature.
}
\label{fig:detuning_multiple}
\end{figure}

The influence of detuning $\tau_\text{A}\neq\tau_\text{B}$ will be quantified in
the case of Eq. (\ref{eq:phi_simple}). The PGF in Eq. (\ref{eq:g_coding}) can be
used to investigate several probabilities of interest for phase-time coding.
Here, the probability that each party detects exactly one photon
\begin{align}
p(n_\text{A}=1;n_\text{B}=1) = \partial_\text{A}\partial_\text{B} g(0,0)
\end{align}
is addressed, being the central figure
of merit for phase-time coding, as well as the detection of multiple photon
pairs
\begin{align}
p(n_\text{A}\geq 2;n_\text{B}\geq 2) =& 1 +g(0,0)+\partial_\text{A}g(0,0)+
\partial_\text{B}g(0,0)\notag\\&+\partial_\text{A}\partial_\text{B}g(0,0)
-g(0,1)-g(1,0)\notag\\&-\partial_\text{A}g(0,1)-\partial_\text{B}g(1,0)
\end{align}
that must be discarded during key distillation \cite{Gisin_2001}. The
probabilities to detect exactly one photon pair and multiple photon pairs are
depicted in Figs. \ref{fig:detuning_exact1} and \ref{fig:detuning_multiple},
respectively, for different values of $\varphi$ and $\tau_\text{A}-
\tau_\text{B}$. The remaining values are chosen to be $\Delta_t=10\,$ps,
$\delta_t=0.4\,$ps, based on \cite{Stucki_2006}, and $\langle\hat{N}\rangle=1$
for the sake of simplicity. What is evident from this figure is that the
interference becomes the most visible for $\tau_\text{A}=\tau_\text{B}$ and
vanishes for increasing detuning. The probabilities converge to the constant
probability pertaining to $\varphi=\pi/2$, culminating in the complete
disappearance of Franson interference at a detuning of $\delta_t$. Figure
\ref{fig:detuning_exact1} reveals some very interesting conclusions. It can be
inferred that linear dependence dominates the probabilities for increasing
detuning, which is attributed to the convolution of two rectangle functions
resulting in a triangle function. Close to the limit detuning $\delta_t$, the
probabilities indeed branch out linearly and are symmetrically distributed
around the constant limit curve at $\varphi=\pi/2$. Counterintuitively, however,
the curves pertaining to different values of $\varphi$ exhibit a modest
curvature, causing a substantial asymmetry compared to the constant limit curve
at $\varphi=\pi/2$, which represents a non-trivial observation. Figure
\ref{fig:detuning_multiple} depicts lower probabilities and a similar dependence
on the detuning and $\varphi$. The curvature of the curves is slighter, which
leads to a more symmetric distribution of the probabilities w.r.t. to the
constant limit curve at $\varphi=\pi/2$ than in Fig. \ref{fig:detuning_exact1}.
The detection of multiple photon pairs is also most likely for constructive
interference since in that case the photon numbers are most strongly
correlated.

\begin{figure}
\centering
\includegraphics[width=0.45\textwidth]{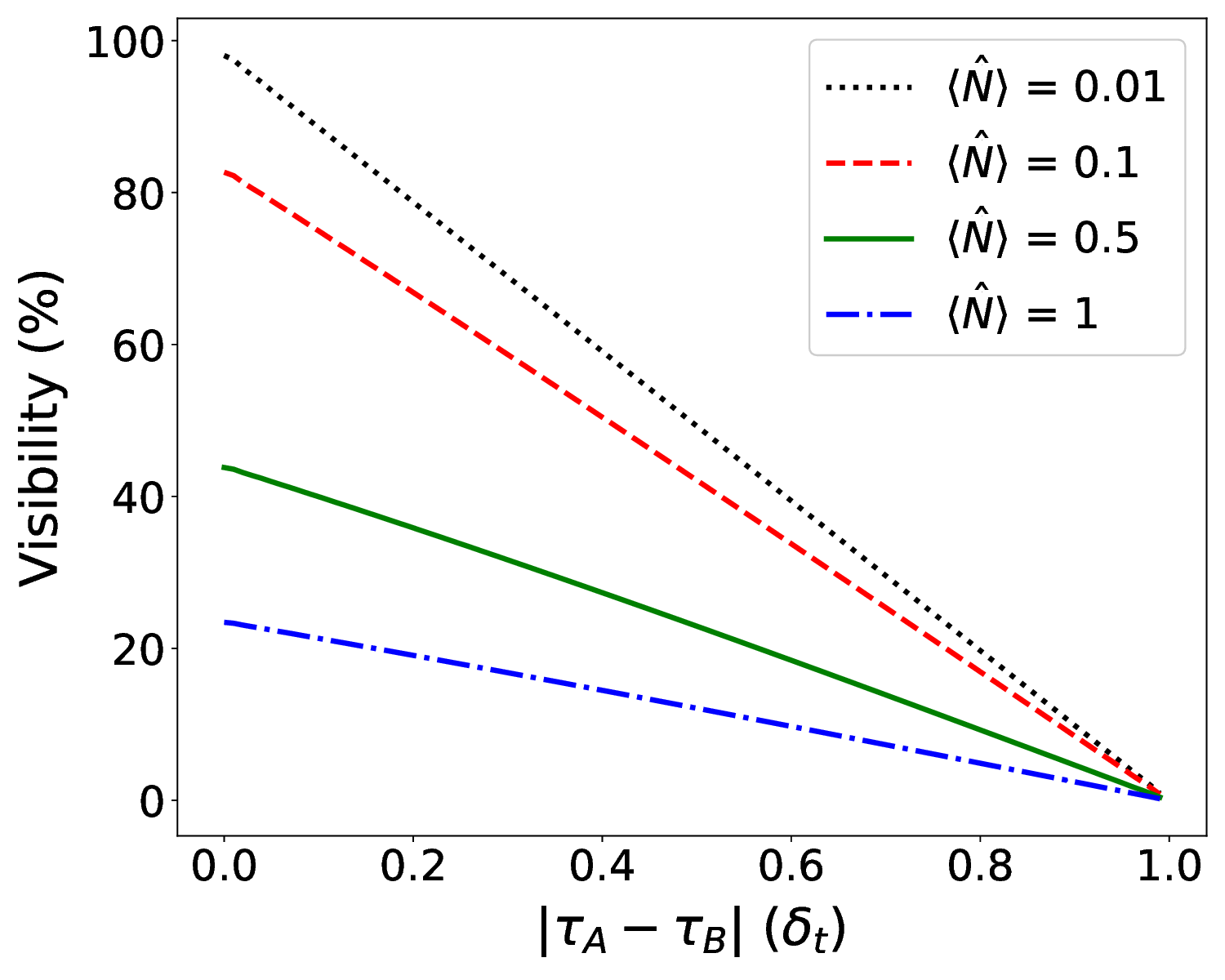}
\caption{Visibility in dependence on the detuning $|\tau_\text{A}-
\tau_\text{B}|$ and the mean photon number $\langle\hat{N}\rangle$. The
visibility increases for lower mean photon number and decreases for higher
detuning, vanishing at $|\tau_\text{A}-\tau_\text{B}|\geq\delta_t$. The curves
are dominated by linear dependence and exhibit a slight curvature, which
decreases for lower mean photon number.
}
\label{fig:visibility}
\end{figure}

These results can be used to estimate an upper bound of the detuning, such that
the security of the key is not compromised. The quality of interference can be
characterized by the visibility:
\begin{align}
V =\frac
{p\vert_{\varphi=0}-p\vert_{\varphi=\pi}}
{p\vert_{\varphi=0}+p\vert_{\varphi=\pi}}.
\end{align}
For practical QKD installations, the visibility typically amounts to at least
93$\%$ \cite{Papapanos_2020}. It is well known that the visibility of Franson
interference substantially depends on the mean photon number, mainly attributed
to multiple photon pairs. Hence, Fig. \ref{fig:visibility} displays the
visibility for different values of the detuning and the mean photon number. As
one might have anticipated, the visibility decreases for higher detuning,
vanishing at $|\tau_\text{A}-\tau_\text{B}|\geq\delta_t$, and increases for
lower mean photon number. Once again, the dependence on the detuning is mainly
linear and exhibits a slight curvature, which decreases for lower mean photon
number. To ensure the security of the key, it can be inferred that the detuning
should not drop below 6.9$\%$ in the case of $\langle\hat{N}\rangle=0.01$. This
estimation represents a powerful tool to assess the impact of the detuning on
Franson interference. Moreover, the mean photon number should not exceed a
certain limit, even in the case of no detuning. This limit turns out to be
0.037.

\section{Conclusion}\label{s:conclusion}
In summary, a theoretical approach was presented, which provides full
time-dependent counting statistics of photon pairs generated by spontaneous
parametric processes in terms of efficiently computable formulas. The regime of
validity is given by an arbitrarily high degree of entanglement and a low degree
if the mean photon number is bounded. The counting statistics were derived for
three types of time intervals, which are classified according to their widths.
The joint amplitude could be chosen arbitrarily and, as communication channels,
free space and fiber propagation were investigated.

Apart from direct detection of the entangled photon pairs, general setups and
external influences were discussed. It was shown that the approach can be easily
generalized and a general procedure to modify the derived formulas was
presented. As an example of the utility of the approach, it was applied to
phase-time coding and full description of Franson interference for increasing
detuning of the MZIs was presented. Up to now, this could not be achieved since
previous approaches did either not include all physically contributing
information or not provide an efficient computation for arbitrarily high
entanglement. The detection probabilities were derived to investigate this
impact, revealing interesting conclusions on how Franson interference is
affected by the detuning. On this basis, an acceptable range of the detuning was
estimated for practical QKD installations.

The present paper is likely to motivate future studies as its regime of validity
and the description of intermediate interval widths can be extended by
investigating the approximations of the presented approach. Several applications
of the present work can be envisaged. First, many kinds of experimental setups
and external influences could be modeled and the analytic results could be
compared to measurement outcomes. Secondly, the influence of noncollinear
geometries and fiber coupling to counting statistics is also of appreciable
interest. Finally, the assessment of the communication security of a practical
QKD installation is clearly called for.

\appendix
\section{General counting statistics}\label{app:statistics}
The general counting statistics in Eq. (\ref{eq:g_general}) are derived for one
party and one spatial mode, which can be easily extended to the general result.
The proof is inspired by \cite{Takeoka_2015}, where a similar result has been
derived for squeezed states with finite modes and the detection of zero or
non-zero photons, which coincides here to the special case of $y=0$. The
characteristic function of an operator $\hat{A}$ is given by
\begin{align}
\chi_{\hat{A}}(\boldsymbol x,\boldsymbol p) \coloneqq  \text{Tr}\left[ \hat{A} 
\, e^{-i ( \boldsymbol x\cdot\hat{\boldsymbol x} + \boldsymbol p\cdot
\hat{\boldsymbol p} ) } \right].
\end{align}
For squeezed states $\hat{\rho}$, this can be written as \cite{Olivares_2012}
\begin{align}
\chi_{\hat{\rho}}(\boldsymbol x,\boldsymbol p)
= \exp\left[-\frac{1}{2} \begin{pmatrix}
\boldsymbol x \\ \boldsymbol p
\end{pmatrix}^{\text{T}}  \sigma \begin{pmatrix}
\boldsymbol x \\ \boldsymbol p
\end{pmatrix}
\right],
\end{align}
where $\sigma$ denotes the covariance matrix. To address $y^{\hat{N}_I}$, the
interval $I$ is partitioned into $m$ subintervals. Mehler's formula
\cite{Erdelyi_1995} gives
\begin{align}
\notag
\chi_{y^{\hat{N}_I}}(\boldsymbol x,\boldsymbol p)
=& (1-y)^{-m} \left[\prod_{t\notin I} 2\pi \delta(x_t)\delta(p_t)\right]
\\&\times\exp\left[-\frac{1+y}{4(1-y)} \begin{pmatrix}
\boldsymbol x \\ \boldsymbol p
\end{pmatrix}^{\text{T}}  P_I \begin{pmatrix}
\boldsymbol x \\ \boldsymbol p
\end{pmatrix}\right] ,
\end{align}
\\
where $P_I$ denotes the projection onto $I$. Hence, all information of the
quantum state at $t\notin I$ will be eliminated. Applying the trace rule
\cite{Ferraro_2005}
\begin{align}
\text{Tr}[\hat{A}\hat{B}] &= (2\pi)^{-m} \int_{\mathbb{R}^m} \text{d}^m x\, 
\text{d}^m p\, \chi_{\hat{A}}(\boldsymbol x,\boldsymbol p) \,\chi_{\hat{B}}
(-\boldsymbol x,-\boldsymbol p)
\end{align}
for $m$ dimensions to $\langle y^{\hat{N}_I} \rangle = \text{Tr}[\hat{\rho} 
\,y^{\hat{N}_I}] $ gives
\begin{widetext}
\begin{align}
\notag
\left\langle y^{\hat{N}_I} \right\rangle
=& [2\pi(1-y)]^{-m} \int_{\mathbb{R}^m} \text{d}^m x\, \text{d}^m p \,
\exp\left\lbrace-\frac{1}{2} \begin{pmatrix} 
\boldsymbol x \\ \boldsymbol p \end{pmatrix}^{\text{T}}
\left[ P_I\sigma P_I +\frac{1+y}{2(1-y)}  P_I \right] \begin{pmatrix} 
\boldsymbol x \\ \boldsymbol p \end{pmatrix}\right\rbrace
\\\notag
=& (2\pi)^{-m} \int_{\mathbb{R}^m} \text{d}^m x\, \text{d}^m p \,
\exp\left\lbrace-\frac{1}{2} \begin{pmatrix}
\boldsymbol x \\ \boldsymbol p \end{pmatrix}^{\text{T}} 
\left[ (1-y)P_I\sigma P_I +\frac{1+y}{2}  P_I \right] \begin{pmatrix}
\boldsymbol x \\ \boldsymbol p \end{pmatrix}\right\rbrace
\\\notag
=& \det\left[ (1-y)P_I\sigma P_I
+\frac{1+y}{2} P_I \right]^{-\frac{1}{2}}
\\
=& \det\left[ \mathbb{I}+
(1-y)P_I\left(\sigma-\frac{1}{2}\mathbb{I} \right) 
\right]^{-\frac{1}{2}}.
\end{align}
\end{widetext}
The limit $m\rightarrow\infty$ yields the result, where $\sigma$ becomes an
operator and det denotes a Fredholm determinant. 

\section{Proof of Theorem \ref{theorem}}\label{app:theorems}
\begin{proof}
Here, the result concerning $[f(r)](\boldsymbol{x},\boldsymbol{x}')$ is derived.
The idea of the proof can be easily generalized to obtain the remaining results.
First of all, it can be shown that
\begin{align}\label{eq:r(t,t')}
\notag
r(\boldsymbol{x},\boldsymbol{x}')=
&\det\left(\frac{\delta_{\boldsymbol{x}}^{-1}}{\sqrt{2\pi}}\right)
\\
&\times\mathcal{F}_{\boldsymbol{\kappa}}\left[\left\vert\phi
\left(\frac{\Delta_{\boldsymbol{x}}^{-1}}{2}(\boldsymbol{x}+\boldsymbol{x}'),
\boldsymbol{\kappa}\right)\right\vert \right]\bm{(}\delta^{-1}_{\boldsymbol{x}}
(\boldsymbol{x}-\boldsymbol{x}')\bm{)}
\end{align}
satisfies $r^2 = \psi\psi^\dagger$, where the kernel of the integral operator
$\psi$ is given by the JA $\psi(\boldsymbol{x}_\text{A},
\boldsymbol{x}_\text{B})$, defined in Eq. (\ref{eq:ja}). Considering powers
$1\leq n\ll 2/\Vert\Delta_{\boldsymbol{x}}^{-1}\delta_{\boldsymbol{x}}
\Vert_\sigma$ of $r$ as
\begin{align}\label{eq:r^n(t,t')}
\left[r^n \right](\boldsymbol{x},\boldsymbol{x}')
= \int_{\mathbb{R}^3} \text{d}^3 x_1\ldots\text{d}^3 x_{n-1} \,
r(\boldsymbol{x},\boldsymbol{x}_1)\ldots r(\boldsymbol{x}_{n-1},\boldsymbol{x}')
\end{align}
leads to the heart of the proof. The narrow anti-diagonal dependence of
$r(\boldsymbol{x}_j,\boldsymbol{x}_{j+1})$ guarantees the estimation
\begin{align}
\vert \delta_{\boldsymbol{x}}^{-1} (\boldsymbol{x}_j-\boldsymbol{x}_{j+1}) \vert 
\leq 1
\end{align}
for any $j=0\ldots n-1$, including $\boldsymbol{x}_0\coloneqq\boldsymbol{x}$ and
$\boldsymbol{x}_n\coloneqq\boldsymbol{x}'$. Iterative application of this
inequality gives
\begin{align}
\vert \delta_{\boldsymbol{x}}^{-1} (\boldsymbol{x}_j-\boldsymbol{x}) \vert \leq 
j\,,\,
\vert \delta_{\boldsymbol{x}}^{-1} (\boldsymbol{x}_j-\boldsymbol{x}') \vert \leq 
n-j.
\end{align}
Inserting Eq. (\ref{eq:r(t,t')}) in Eq. (\ref{eq:r^n(t,t')}) suggests the
approximation
\begin{align}
\phi\left(\frac{\Delta_{\boldsymbol{x}}^{-1}}{2}(\boldsymbol{x}_j+
\boldsymbol{x}_{j+1}),\boldsymbol{\kappa}\right)
\approx
\phi\left(\frac{\Delta_{\boldsymbol{x}}^{-1}}{2}(\boldsymbol{x}+
\boldsymbol{x}'),\boldsymbol{\kappa}\right)
\end{align}
since $\phi(\boldsymbol{\chi},\boldsymbol{\kappa})$ has normalized widths and
\begin{align}
\notag
&\left| \frac{\Delta_{\boldsymbol{x}}^{-1}}{2}(\boldsymbol{x}_j+
\boldsymbol{x}_{j+1})-\frac{\Delta_{\boldsymbol{x}}^{-1}}{2}(\boldsymbol{x}+
\boldsymbol{x}') \right|
\\\notag
\leq& \frac{\Vert\Delta_{\boldsymbol{x}}^{-1}\delta_{\boldsymbol{x}}
\Vert_\sigma}{2}(\vert \delta_{\boldsymbol{x}}^{-1}(\boldsymbol{x}_j-
\boldsymbol{x}) \vert + \vert \delta_{\boldsymbol{x}}^{-1}(\boldsymbol{x}_{j+1}-
\boldsymbol{x}') \vert)
\\
\leq& \frac{\Vert\Delta_{\boldsymbol{x}}^{-1}\delta_{\boldsymbol{x}}
\Vert_\sigma}{2}(n-1) \ll 1
\end{align}
using $n\ll 2/\Vert\Delta_{\boldsymbol{x}}^{-1}\delta_{\boldsymbol{x}}
\Vert_\sigma$. This yields a convolution of $n$ factors, such that applying the
convolution theorem \cite{Arfken_2005} gives
\begin{widetext}
\begin{align}
\notag
\left[r^n \right](\boldsymbol{x},\boldsymbol{x}')
=& \det\left(\frac{\delta_{\boldsymbol{x}}^{-1}}{\sqrt{2\pi}}\right)^{n}
\int_{\mathbb{R}^3} \text{d}^3 x_1\ldots\text{d}^3 x_{n-1} \,
\\\notag&
\mathcal{F}_{\boldsymbol{\kappa}}\left[\left|\phi
\left(\frac{\Delta_{\boldsymbol{x}}^{-1}}{2}(\boldsymbol{x}+\boldsymbol{x}_1),
\boldsymbol{\kappa}\right)\right| \right]\bm{(}\delta^{-1}_{\boldsymbol{x}}
(\boldsymbol{x}- \boldsymbol{x}_1)\bm{)}\ldots\,
\mathcal{F}_{\boldsymbol{\kappa}}\left[\left|\phi
\left(\frac{\Delta_{\boldsymbol{x}}^{-1}}{2}(\boldsymbol{x}_{n-1}+
\boldsymbol{x}'),\boldsymbol{\kappa}\right)\right| \right]\bm{(}
\delta^{-1}_{\boldsymbol{x}}(\boldsymbol{x}_{n-1}-\boldsymbol{x}')\bm{)}
\\\notag\approx& \, 
\det\left(\frac{\delta_{\boldsymbol{x}}^{-1}}{\sqrt{2\pi}^n}\right)
\overbrace{
\mathcal{F}_{\boldsymbol{\kappa}}\left[\left\vert\phi
\left(\frac{\Delta_{\boldsymbol{x}}^{-1}}{2}(\boldsymbol{x}+\boldsymbol{x}'),
\boldsymbol{\kappa}\right)\right\vert \right]
\ast\ldots\ast
\mathcal{F}_{\boldsymbol{\kappa}}\left[\left\vert\phi
\left(\frac{\Delta_{\boldsymbol{x}}^{-1}}{2}(\boldsymbol{x}+\boldsymbol{x}'),
\boldsymbol{\kappa}\right)\right\vert \right]
}^{n}
\bm{(}\delta^{-1}_{\boldsymbol{x}}(\boldsymbol{x}-\boldsymbol{x}')\bm{)}
\\=& \, 
\det\left(\frac{\delta_{\boldsymbol{x}}^{-1}}{\sqrt{2\pi}}\right)
\mathcal{F}_{\boldsymbol{\kappa}}\left[\left\vert\phi
\left(\frac{\Delta_{\boldsymbol{x}}^{-1}}{2}(\boldsymbol{x}+\boldsymbol{x}'),
\boldsymbol{\kappa}\right)\right\vert^n \right]
\bm{(}\delta^{-1}_{\boldsymbol{x}}(\boldsymbol{x}-\boldsymbol{x}')\bm{)}.
\end{align}
\end{widetext}
This result can be extended to a general analytic function $f(x)$ by expanding
it as a power series. Since the proven relation, however, only holds for
$n\ll 2/\Vert\Delta_{\boldsymbol{x}}^{-1}\delta_{\boldsymbol{x}}\Vert_\sigma$,
higher-order terms have to be negligible. The argument of $f(x)$ can be
estimated by the maximal value $|x|\leq\Vert\phi\Vert_\infty$. To obtain
$x_\text{max}$, the mean photon number $\langle\hat{N}\rangle$ can be introduced
using Eqs. (\ref{eq:quadr_exact}) and (\ref{eq:N_estim}), which hold for any
choice of $\Delta_{\boldsymbol{x}}$ and $\delta_{\boldsymbol{x}}$.
\end{proof}

\end{document}